\documentclass[10pt,preprint]{aastex}
\usepackage{psfig}

\begin{document}

\newcommand \gam{\gamma}
\newcommand \bb{{\mit \mathbf B}}
\newcommand \be{{\mit \mathbf E}}
\newcommand \bk{{\mit \mathbf k}}
\newcommand \veperp{\varepsilon_\perp}
\newcommand \epsperp{\epsilon_\perp}
\newcommand \pperp{{\mit \mathbf p_\perp}}
\newcommand \lap{\lesssim}
\newcommand \gap{\gtrsim}
\newcommand \thm{\theta_{\rm max}}
\newcommand \eps{\epsilon}
\newcommand \om{\omega}
\newcommand \bom{\mathbf \Omega}
\newcommand \vep{\varepsilon}
\newcommand \ppar{{p_\parallel}}
\newcommand \bc{B_{cr}}
\newcommand \kap{\kappa}
\newcommand \gcm{\gamma_{\rm CM}}
\newcommand \betacm{\beta_{\rm CM}}
\newcommand \pave{\bar{\ppar}}
\newcommand \gave{\bar{\gamma}}
\newcommand \gray{{$\gamma$-ray }}
\newcommand \grys{{$\gamma$-rays }}
\newcommand \cd{{$\cdots$}}
\newcommand \etal{{\it et~al. }}
\newcommand \eg{{\it e.g. }}
\newcommand \ApSS{{Ap.~Space Sci.\/}}

\title{Pair Creation in the Pulsar Magnetosphere}

\author{Paul N. Arendt, Jr.}
\email{parendt@aoc.nrao.edu}
\author{Jean A. Eilek}
\email{jeilek@aoc.nrao.edu}
\affil{Department of Physics, New Mexico Tech, Socorro, NM 87801}

\begin{abstract}

We present numerical simulations of the electron-positron
plasma creation process in a simple neutron star magnetosphere.
We have developed a set of cascade `kernels', which represent the
endpoint of the pair cascades resulting from monoenergetic photon
seeds.  We explore two popular models by convolving these kernels  
with the seed photon distributions produced by curvature radiation and
by inverse Compton scattering.  
We find that the pair plasma in either case is
well-described in its rest frame by a relativistic Maxwellian
distribution with temperature near $mc^2/k_B$.  We present
cascade multiplicities and efficiencies for a range of seed particle
energies and stellar magnetic fields.  We find that the efficiencies
and multiplicities of pair creation are often lower than has been
assumed in previous work. 

\end{abstract}


\section{INTRODUCTION}

An electron-positron pair plasma is a key ingredient in most models
of pulsar radio 
emission.  This plasma is assumed to come from a pair cascade which
occurs in the open field line region close to the star's magnetic
axis.  In this region,  rotation-induced electric fields pull charged 
particles from the polar cap and accelerate them to relativistic
energies ($\gam \lap 10^7$).  These particles  radiate `seed'
\gray photons, 
either by curvature emission (Sturrock 1971; Ruderman \& Sutherland
1975; Arons 1983), or by inverse Compton scattering of ambient
thermal photons (Bussard, Alexander, \& M\'esz\'aros 1986; Daugherty
\& Harding 1986; Sturner \& Dermer 1994).  In
the strong magnetic fields of pulsars, these
primary photons are susceptible to magnetic one-photon
electron-positron pair creation (Tsai \& Erber 1974).  The
newly formed leptons in turn radiate `secondary' photons, most
commonly through synchrotron radiation (Harding \& Preece 1987). The
secondary photons may be capable of further pair production.  As
Sturrock (1971) first pointed out, this cycle of energetic photon
emission and further pair creation continues, forming a `pair
production avalanche', which ends only when all remaining photons are
transparent to pair creation.  

Whether this pair plasma forms or not, and its properties when it
does, are crucial issues in models of the radio emission region.  
The plasma properties determine the possible wavemodes which the
plasma can support.  Excitations of these
wavemodes ultimately become the radio emission we observe (after
coupling to electromagnetic modes and escaping the
magnetosphere). Propagation of these modes through the pair plasma may
leave its signature 
on the observed signal (through dispersive effects).  In addition, the
plasma itself may be the source of the excitation 
of the waves which lead to radio emission (if there is
free energy available in the plasma distribution at birth
to drive instabilities). 

The need for a quantitative understanding of the pair plasma 
distribution function (DF) has
been apparent in the literature for some time.  Existing calculations
of plasma wavemodes and propagation have either assumed analytically
convenient DFs, without physical justification, or have tried to
quantify a plausible heuristic DF introduced by Arons (1981). 
(Examples include  Buschauer \& Benford 1975, Arons \& Barnard 1986,
Beskin, Gurevich, \& Istomin 1988; Kazbegi, Machabeli, \& Melikidze
1991; Weatherall 1994, Lyubarskii 1996;  
Gedalin, Melrose, \& Gruman 1998; Lyutikov, Blandford, \& Machabeli
1999).  It is therefore critical to
determine the properties of the plasma created by a pair-production
avalanche in the pulsar magnetosphere.  This is the primary focus of
our paper. 

In addition to leptons, the pair cascade can also produce high-energy
photons. A small number of pulsars exhibit such high-frequency
emission.  Some authors ({\it e.g.,} Romani 1996;  Horotani \& Shibata
1999) believe that this emission comes from a high-altitude `outer
gap' active region, others ({\it e.g.}, Rudak \& Dyks 1999;  Zhang \&
Harding 2000) believe that this emission originates in the polar cap
region.  We therefore include photon spectra as a secondary
focus in the cascade models we present in this paper. 

\subsection{The Setting: the Polar Flux Tube}

To place our calculation into a larger context,
we briefly summarize the standard model of the pulsar magnetosphere.
(We follow, for instance, Ruderman \& Sutherland 1975, Arons 1992,
or Melrose 1992, 1995).

Soon after pulsars were discovered, Goldreich \& Julian
(1969) pointed out that a pulsar magnetosphere would
not be empty.  A rotating magnetized neutron star in vacuum
generates electric fields strong enough to overcome the star's binding
energy for electrons (and possibly light ions), contradicting the
vacuum assumption.  Most of the inner
magnetosphere is now assumed to corotate with a `force-free'
electric field $\be_{ff} = - c^{-1} (\bom \times {\mathbf r})
\times \bb$, filled with the
corotational (Goldreich-Julian) charge density $\rho_{GJ} = 
(4 \pi)^{-1} \nabla \cdot \be_{ff}$,
where $\bom$ is the angular velocity of
the star.  The exception is on those field lines
which extend beyond the `light cylinder' (where the corotational
speed is equal to $c$), defining the `polar flux tube' (Arons 1983).
Radio emission is thought to originate in the plasma within this polar
flux tube.   

The polar flux tube's active properties are due to
the extension of its $\bb$ field lines beyond the light cylinder.
Charged particles which stream outward along these lines are unable
to maintain a corotational force-free state along the entire field
line (even within the light cylinder).
(In the `closed' magnetosphere, where $\bb$ lines
do not cross the light cylinder, $\bb$ lines become electric
equipotentials.)  Although individual particles in the flux
tube may escape to form a stellar wind,
polar currents are assumed to cross field
lines somewhere near the light cylinder and complete a circuit
back to the star to preserve its overall neutrality.  

The details of how these polar currents return to the star are
unknown.  This is unfortunate, because the global current structure is
crucial for determining the accelerating electric fields, and thus the
photon seeds for the pair production cascade (if it occurs).  If a
cascade occurs, the pair plasma may allow
the premature `shorting out' of the accelerating potential (\eg Arons \&
Scharlemann 1979; Shibata, Miyazaki, \& Takahara 1998).  This in turn
modifies the conditions assumed to seed the cascade in the first place.

\subsection{Modeling The Pair Cascade}

The pair cascade takes place within this setting.  We have already
noted the complexity of a fully self-consistent solution.  Lacking
this, important factors for the cascade development are
uncertain.  What is the energy of the primary beam charges?  By what
means do these primary charges produce seed \gray photons? 
In contrast, the microphysics underlying the cascade is well-known.
We know precisely the differential QED cross sections for lepton and
photon production: the local $\bb$ field, photon energy and impact
angle determine the outcome. 

Daugherty and Harding (1982; `DH82' hereafter) modeled the cascade
numerically.  Their simulations began with a single particle streaming
out along the $\bb$ field, whose curvature radiation began the
cascade. They
studied the \gray photon distribution produced by the cascade, with
passing reference to the properties of the underlying plasma. 

Our focus is different.  Because we are especially concerned with the
properties of the pair plasma, we wanted to extend previous work to
determine those properties, and their dependence on
magnetospheric parameters (magnetic field and primary beam energy).
Thus, we set up our calculation
to determine both the momentum distribution function (DF) and the 
density of the pair plasma relative to that of the primary beam.

Because we were concerned about the uncertainties implicit in global
magnetosphere models, we took a new approach to the cascade.  
We wrote a code which follows, in time and space, the cascade induced
by a monoenergetic population of photons. We include pair production
by the photons and synchrotron radiation by the leptons.  Each run
terminates when all particle and photon production ceases.
At the end of each run, we store  the photon and lepton distributions
(typically about $10^5$ particles per run, binned into 
50-100 momentum bins, depending on the number of particles available).
We treat each such run as a `kernel'.  The
final cascade is then formed as the composite of many such runs, each
weighted by the relative distribution of seed photons of that
particular energy.   In this paper we form composite cascades assuming
the seed photons come from curvature radiation, or from inverse
Compton scattering of ambient X-ray photons. 

In the remainder of this paper, we describe  our code (\S 2; with some
details in the Appendix);  give a qualitative overview of the pair
cascade process (\S 2), and   present the results of our
parameter-space survey (\S 3).
Our primary results are presented in \S 4, where we form
composite cascades, based on photon seeds from curvature radiation and
inverse Compton scattering.  We close in \S 5 with a summary and 
some final comments. The reader interested only in final plasma or photon
distributions might skip the details and jump to \S \S 4 and 5.

\section{OUR KERNELS: MONOENERGETIC SEEDS}

\subsection{Structure of the Calculation}

Our simulations began with photons injected over a polar cap, at the
stellar surface (actual seed photon formation at a finite but low altitude
will encounter a similar physical environment).  The magnetic field
geometry is assumed to be
dipolar.  The photons are monoenergetic (to within one percent) with
energy $\vep$.  The
initial locations are uniformly scattered in area on the stellar
surface over the entire polar cap (a circle centered on the magnetic
pole, of radius  $r_* \sqrt{\Omega r_*/c},$ where $r_*$ is the stellar
radius).  The photons are distributed
uniformly in a cone of half-angle $\thm$ relative to the local $\bb$
direction.   

The major input parameters for each run are: the 
magnetic field strength, $B_*$ (which we describe in terms of the
field value at the magnetic pole, at the altitude corresponding to the
photon seed emission), the energy $\vep$ of the primary photons seeding
the cascade, and their maximum initial angle $\thm$ (to which we
refer using $\mu = 1 - \cos \thm$).  The range of
these parameters covered by our complete set of
monoenergetically-seeded cascades was determined partly by the
demands of the specific model cascade seed mechanisms of \S
4, and partly by our performing a systematic exploration over a
wide range of these parameters (to determine the influence of each
parameter on the cascade process).  Our complete set of simulations
covered initial photon energies $\vep \equiv \hbar \om/mc^2$ (where
$m$ is the electron mass)
from $50$ to $10^7$, initial field values $B_* = 10^{11},
10^{12}$, and $10^{13}$ G, and cosine complements
$\mu$ from $0$ up to $0.1$.  A subset of these
results is presented in this section, with parameters chosen from
a representative sample of the range covered.  The smallest opening
angles
shown here correspond to $\mu = 10^{-12}$; we found that lowering
$\mu$ beyond this usually had no further effect on the
cascade (the
exceptions to this were the $B_* = 10^{13}$ runs at the largest
initial photon energies $\vep$ that we investigated; there, smaller
$\mu$ lowered the cascade multiplicities somewhat).
Angles larger than the Lorentz
beaming cone, $\thm \sim 1/\gamma$ (where $\gamma$ is the Lorentz
factor of the beam particles emitting the initial photons), are
not believed to be relevant in present-day models of the pulsar pair
cascade (which all rely upon a high-energy particle beam to emit
or scatter the seed photons).
We nevertheless included some larger angles
in our monoenergetic runs to investigate
the variation of cascade possibilities (but did not use them in our
composite models, presented in \S 4, which were all from a 
simulated particle
beam).  We present them here in case
future models include primary photons at high angles (such as
might be expected from energetic return currents impacting the star
when inverse Compton braking is negligible).

Our code follows the position and momentum of each photon as it
propagates away from the 
star.  To determine when a photon pair creates, we also track the optical
depth, $\tau$, for each photon.  For photons, $\tau$ is the fraction of a
mean free path traversed before pair production occurs, so that
$\tau = 1$ represents a $1/e$ probability that it has not
disappeared and created a pair.  
We split the simulation into discrete timesteps (with duration
chosen so that the optical depth of no photon
 increases by more than $\delta \tau = 0.2$ per timestep). 
The photons thus begin by propagating outward unattenuated,
until some of them exceed $\tau = 1$ and create
electron-positron pairs.

The local conditions for each photon (field strength, particle energy
and angle) determine how much its $\tau$ is incremented at each timestep.
For photons of energy $\vep$,
the main relevant parameter is $\chi$,
defined as (Daugherty \& Harding 1983)
$$
 \chi = {{\vep} \over { 2 }} B' \sin \theta, 
\eqno(1)
$$
where $\theta$ is the angle between the
photon's direction and $\bb$, and
$B' = B/\bc$ is the dimensionless magnetic field strength,
The `critical' magnetic field strength is
$ \bc \equiv m^2 c^3 /e \hbar = 4.4 \times 10^{13}$ G.
The total attenuation coefficient
$R(\chi)$ (given fully in the Appendix), can be approximated at low
$\chi$ values by
$$ 
R(\chi) \approx 0.23 {{\alpha} \over {\lambda_C}} B' \exp \left( -
{{4} \over {3 \chi} } \right)  \, \, , 
\eqno(2)
$$
where $\alpha$ is the fine-structure constant, and $\lambda_C$
is the Compton wavelength of the electron.

When we exceed $\tau = 1$ for a photon, that photon is destroyed and
an electron-positron pair created in its place.
The properties of the new leptons are
determined probabilistically, using numerical approximations to
the differential attenuation
coefficients for pair production into the possible lepton states.
Expressions for these coefficients are taken from Daugherty \&
Harding (1983), and included in the Appendix.

The possible states of the created pair depend upon the state of
the parent photon, allowed states of charged leptons in strong
magnetic fields, and relevant conservation laws.
The energy levels $E/(mc^2)$ of an electron or positron with momentum
${\mathbf p}/(mc)$
in a uniform $B$
field are given by (Johnson \& Lippman 1949)
$$
E_{n,s,\ppar}^2 = 1 + \ppar^2 + B'(2n \pm s + 1) 
\eqno(3)
$$
where $\ppar = {\mit{\mathbf p}} \cdot \bb/B$ is momentum along $\bb$,
$s$ can be $\pm 1$ and specifies the spin state, and $n$
is a nonnegative integer specifying the orbital quantum number (Landau
level).  The
leptons may be born into an excited Landau level and emit
high-energy synchrotron radiation, which can create further
lepton pairs.  After the rapid synchrotron emission a newly created
lepton emits in these strong fields, it will be in the ground state
($n=0$ and $1 \pm s = 0$), constrained to move along $\bb$.

We also track the position and momentum of each lepton, constraining
its guiding center to move along $\bb$.
To decide when to create a new synchrotron photon, we use a
semi-classical method 
to simulate the lifetime of the parent lepton's current quantum
state, and use this to define a mean free path.  Details are given in
the Appendix.  As with the photons, we
use the fraction of a local mean free path traversed during each
timestep to increment $\tau$.  When $\tau > 1$, we create a
single synchrotron photon.  Keeping track of emission
times is most important in weaker $\bb$ fields ($B \ll \bc$),
as all synchrotron
emission is extremely rapid in strong fields.

For leptons in the strong magnetic fields of pulsars, the quantized
nature of synchrotron emission cannot be ignored.  The energy
difference
between adjacent Landau levels is $mc^2 B'$, so synchrotron
photons can easily exceed $\vep = 2$ in strong fields, and
in weaker fields when the leptons are born with high $\pperp$
(where $\pperp$ is the momentum at right angles to $\bb$,
implictly defined by the Landau level of eq. [3]).
As discussed in Harding \& Preece (1987), the classical synchrotron
spectrum can give erroneous results under these strong $\bb$
conditions: the high-frequency limit of the classical spectrum
can violate conservation of energy, and the low-frequency classical
spectrum can extend below the minimum allowed transition.
As the leptons may lose $\pperp$ either by emission of many
low-energy photons or a few high-energy ones, we choose the
photon energy probabilistically.  This is done using the
differential synchrotron spectrum in Harding \& Preece (1987).
The Appendix gives some details of this spectrum (eq. [A3])
and the code implementation of these processes.

A simulation has reached completion when all photons
satisfying $\vep > 2$ have
essentially zero probability of pair creation, and all leptons have
radiated away all $\pperp$ by synchrotron emission.

Some parameters affect the cascade only very weakly for a given
initial photon injection; for uniformity, we left these fixed.
In all the runs presented here, we set the neutron star
radius at 10 km, with a rotation period of 33 ms.
The angle between the rotation and magnetic dipole axes was set
at $45^\circ$.  Of these fixed parameters, the cascade details
will be most sensitive to the rotation period, due to the more strongly
curved field lines available above the larger polar caps of
faster pulsars.

Note that we have ignored several effects which may have a small
effect on the cascade.  For simplicity, we ignored polarization
and spin dependence in particle creation processes, and only
used cross-sections which have been averaged over polarization
and spin.  We ignored general relativistic effects,
such as frame dragging and gravitational refraction.
Frame dragging leads to higher beam energies (through making
accelerating electric fields stronger), but beam energy is
just a parameter in our (composite) simulations anyway.
We did not include any electric fields in these simulations.
The primary effect a nonzero $\be_\parallel$ would have in our
simulations is to cause relative streaming of the leptons of different
charge.  Another effect is that pair production efficiency is boosted
in the presence of an $\be$ field (Daugherty \& Lerche 1975).
Including this effect is locally equivalent to
increasing $B$, a variation we do investigate here.
We ignored the possibility of {\it bound} positronium formation in strong
($ B > \bc $) fields (\eg Usov \& Melrose 1996), the decay of which
may be more  likely to result in photon multiplication than creation
of a (stable) free pair.  We also ignored photon splitting (a closely
related process, differing only in that the intermediate pair is virtual).
Qualitatively, the inclusion of these higher-order processes might be
expected to lower the
pair creation efficiencies in strong fields, and degrade the
final photon spectrum to lower energies.

\subsection{A Typical Monoenergetic Cascade}

The simulation results will be easier to understand if we first sketch
the qualitative features of some typical monoenergetic
runs.  We choose three simulations: all have initial
photon energy $\vep = 10000$ and cosine complement $\mu =
10^{-6}$.  Although this is a larger angle than is allowed by
$1/\gam_b$ beamed emission from a `primary' beam as in most
models, we can illustrate a wider variety of possible cascade behavior
than with runs at only very small angles.  (We again
emphasize that we respected
$1/\gam_b$ beaming in our composite runs of \S 4.)
These illustrative runs differ from each other only in the surface
magnetic field strength $B_*$ at the magnetic pole: $10^{11},
10^{12}$, and $10^{13}$ G, respectively.  

\subsubsection{Particle production}

In figures 1 and 2 we illustrate the factors governing the existence
and development of a typical cascade.  In each figure, pair creation
events are shown in the left panels, and synchrotron photon creation
events in the right panels.  In addition to the spatial development of
the cascade, these two figures illustrate the two factors which govern
pair creation: opacity and photon energy.

\begin{figure}
\centerline{\psfig{figure=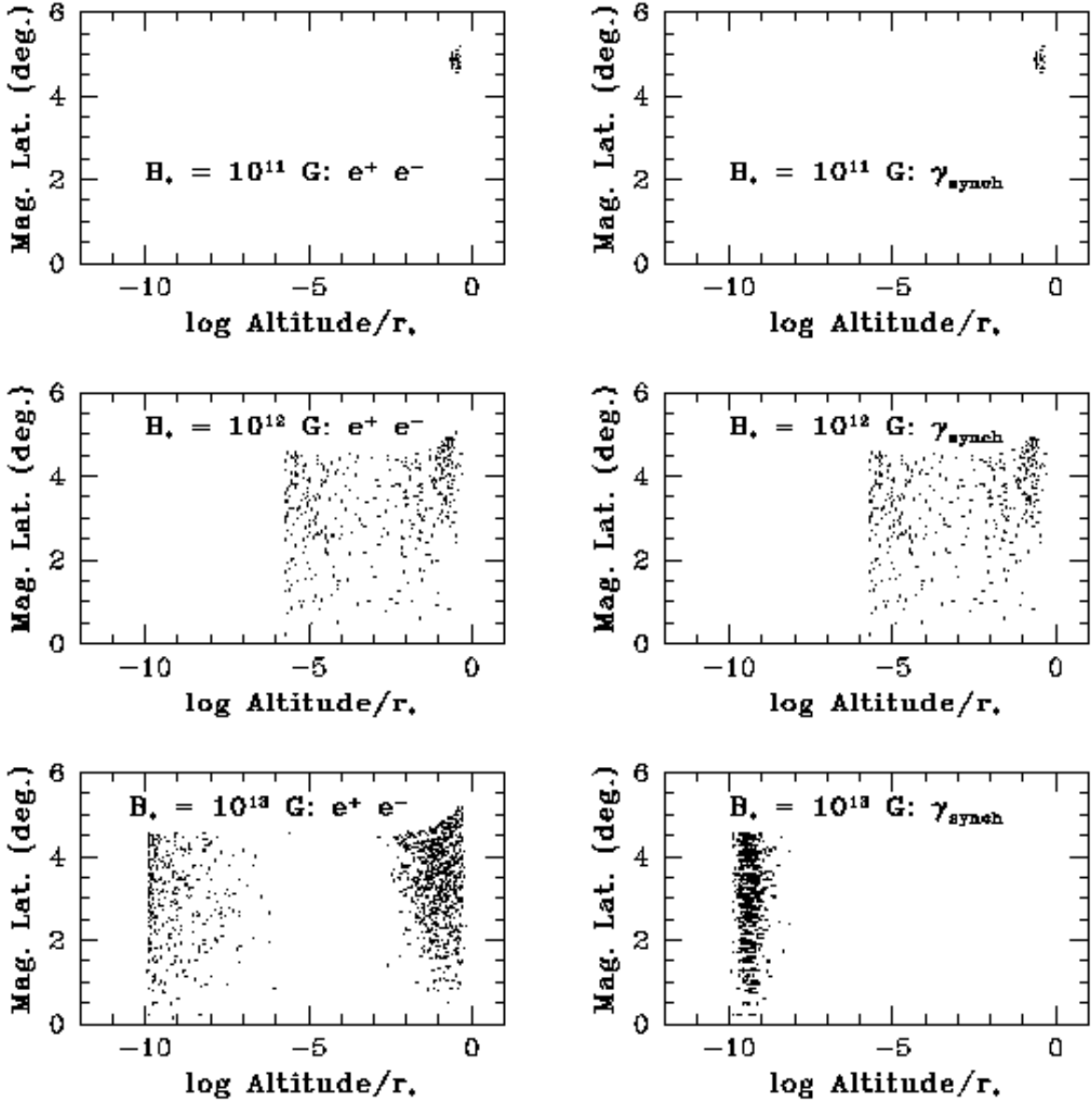,height=17cm,width=17cm}}
\caption[f1.eps] {Locations of Particle Production.
Magnetic latitude $\rho$
is plotted against log altitude (in stellar radii), for
photons of $\vep = 10000$ injected into a cone with angle
parameter ($\mu = 1 - \cos \theta_{\rm max} = 10^{-6}$).
The left column shows sites of lepton pair
creation, and the right column shows sites of energetic ($\eps > 2$)
synchrotron photon emission.  The number of starting photons was
300, 200, and 300 for $B_* = 10^{11}, 10^{12},$ and $10^{13}$ G,
respectively.  At smaller angles (e.g., for $1/\gam_b$ beaming),
the only difference in this diagram is the disappearance of the
low-altitude burst of pairs and synchrotron photons in the
$10^{13}$ G field.}
\end{figure}

\begin{figure}
\centerline{\psfig{figure=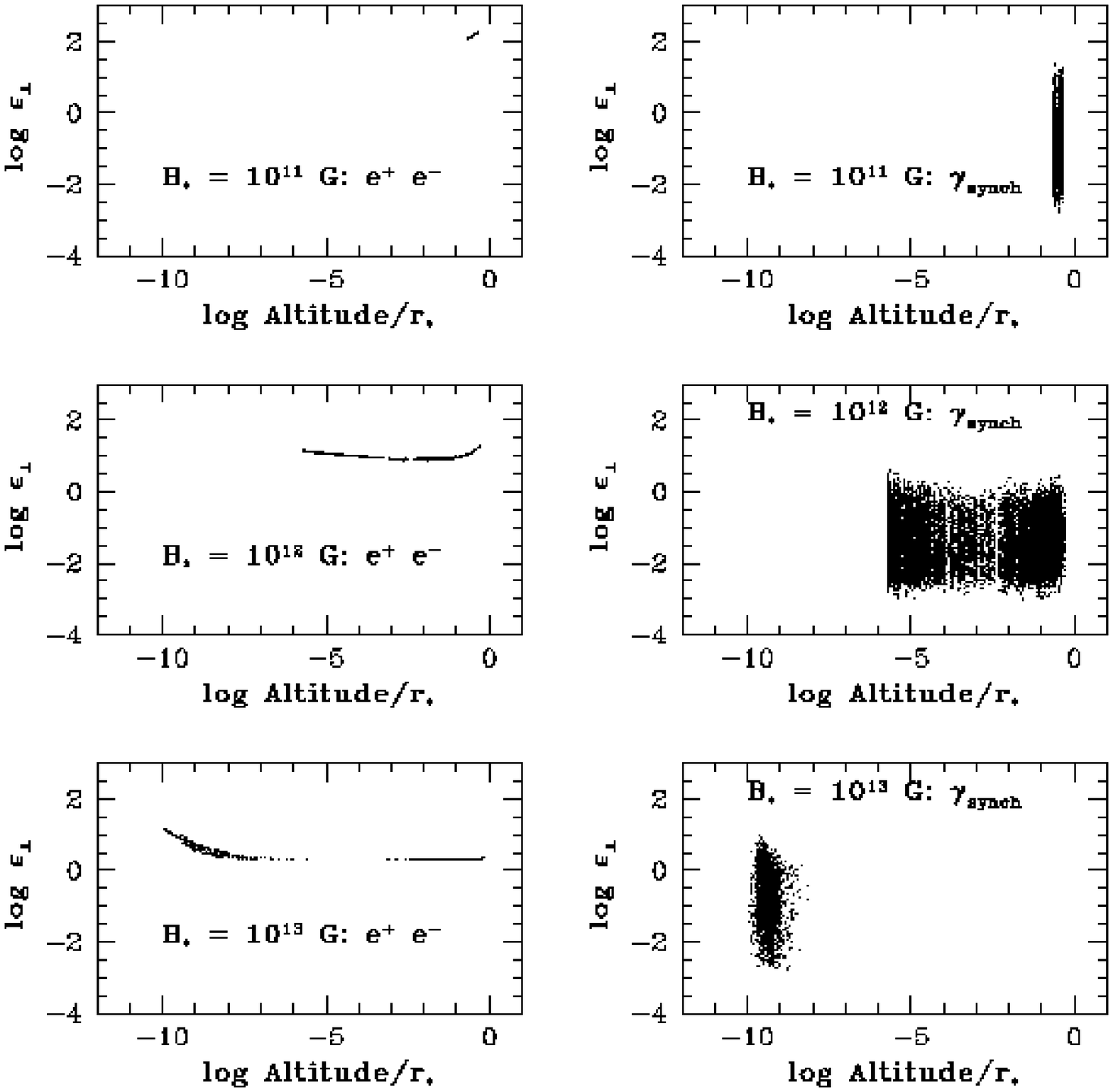,height=17cm,width=17cm}}
\caption[f2.eps] {Photon Energies in Particle Production.
Similar to figure 1,
but showing photon $\eps_\perp$ (energy in a frame where
$\bk \cdot \bb = 0$) versus altitude.
The energy shown is that of the parent photon for the pair production
events, and of the emitted photon for the synchrotron events.  Note
that many synchrotron photons are often emitted right at the sites
of pair creation seen in figure 1.  It is evident that many of
these secondary photons are born with insufficient $\eps_\perp$
($< 2$) to create another pair, unless their angle with $\bb$ increases
significantly.}
\end{figure}

In figure 1 we show the spatial development of the cascade, plotting
magnetic
polar angle against (log) altitude for each pair creation or energetic
$(\eps > 2)$ synchrotron photon creation event. 
(Throughout this work, we use $\vep$ to denote seed photon energies,
and $\eps$ for general  photon energies, including post-cascade.)
Figure 2 is similar to figure 1, but 
instead of magnetic polar angle we show log $\epsperp$ as the ordinate,
where $\epsperp$ refers to the perpendicular frame energy of the
parent photon for pair creation events, and that of the daughter
photon for synchrotron emission events.  
Several features of these figures are worth pointing out. 

Figure 1 shows that the cascade in a $10^{13}$G field has two steps:
a low-altitude burst 
of pair formation and synchrotron emission, followed by a
high-altitude burst of pair creation without further photon creation.
In the two lower-field cases, the cascade has only one burst, in which
both pair and photon creation occur.

Inspection of figure 2 clarifies the two factors which control the
onset of the cascade.  Consider the available energy ($\epsperp$) of
the photon which pair creates.  In all creation events in the lower
field runs, and in the lower-altitude creation events in the
$10^{13}$G run, $\epsperp \gg 2$.  These photons have 
more than enough energy to create a pair.  The leptons created in
these events are born in high Landau levels (they have finite angles 
relative to $\bb$).  
They will therefore create new photons through synchrotron
radiation; the short lifetime of this process is reflected in the
spatial bunching of photon creation events, close to the location of
the original pair event. In contrast, the
high-altitude second burst in the $B = 10^{13}$G field comes from
photons with $\epsperp \simeq 2$.  These creation events 
come from photons  traveling
nearly tangent to $\bb$. Their propagation angles, relative to
$\bb$, increase at higher altitudes due to field line curvature, until
they finally attain $\epsperp \ge 2$, at which point pair creation occurs.
The created leptons are born traveling tangent to $\bb$,
in the ground Landau state, so that they cannot create any further
synchrotron photons.  Thus the high-altitude pair creation burst is
not accompanied by further photon creation.

We can understand this behavior quantitatively.  The high-field,
high-altitude creation events are governed by energetics.  They
illustrate the first condition necessary for pair creation, that the
photon must have at least the rest mass energy measured in the
perpendicular frame:
$$
 \epsperp = \epsilon \sin \theta \ge 2
\eqno(4)
$$

By contrast, the lower-field, lower-altitude creation events are
coverned by opacity.  They can be understood from
the low-$\chi$ behavior of $R(\chi)$, given in equation (2).
Based on our numerical results, and noting that 
$B$ decreases as $(r/r_*)^3$, we can generously estimate that a stellar
radius will be the maximum distance traversed before pair creation.
Setting the mean free path (the inverse of $R$) equal to 10 km gives a
minimum value of $\chi$ realistically necessary for pair production:
$$
\chi_m \approx (23.5 + 0.75 \, \ln B')^{-1} \approx 0.05 \, . 
\eqno(5) 
$$
(This depends only weakly on $B'$, so it can be assumed robust
when changing the magnitude or geometry of the field).
Opacity to pair creation thus  requires $\chi > \chi_m$.  
Using equation (1), we find the second condition necessary for pair
creation: 
$$ 
\epsperp B \gap 0.1 \bc \sim 4 \times 10^{12} \, G \, . 
\eqno(6)
$$
Our runs at $B_* = 10^{11}$ and $10^{12}$ G are thus
opacity limited: the photons pair create when they propagate to
angles so that $\epsperp \gap 0.1 \bc / B$.  This relation describes
the locus of pair creation sites in the two lower $B_*$ diagrams of
figure 2, and the low-altitude creation sites in the $B = 10^{13}$G
run. In contrast, the high-altitude pair creation events in our
$B_* = 10^{13}$ G example are energy limited. 
The photons nearly always satisfy $\chi > \chi_m$, but they cannot
pair create until they also satisfy $\epsperp > 2$. 

In addition to the cascade onset, the cascade termination is worth
comment. 
In all cases (including those not shown), the cascade  finished
before the particles reached one stellar radius above the surface
(or injection point).  Thus,  pair cascade is a rapid process, with
the time from seed photon emission to cascade completion being 
less than $r_* / c \sim 30 \, \mu$sec. If seed photon production
is a sporadic process, then the pair plasma created should be
expected to also exist only sporadically at any given point
in the polar flux tube.  The short completion time is reminiscent of
the temporal, `microstructure' fluctuations seen in pulsar signals
({\it e.g.} Hankins 1996).  We also note that changing the
geometry of the $\bb$ field would have to be significant over
these short length scales, in order to have any noticable
effect on the cascade.  Finally, figure 1 
also shows an enhanced probability of pair production towards the
edge of the polar flux tube.  This is of course due to the higher
field line curvature there, and is reminiscent of `conal' emission (as
defined in Rankin 1983).

\subsubsection{The cascade at other angles}

Most of the features of the examples presented apply to the
cascade at smaller seed photon opening angles.  Figures
1 and 2 lose only the low-altitude creation events in the
$10^{13}$ G field (which includes all synchrotron photons in
that case).  The cascade remains essentially unchanged for
all leptons whose first pair creation burst was
opacity-limited at smaller angles, including down to
$\mu = 0$.

At larger $\mu$, most of the initial pair creation events
are opacity-limited.  Copious synchrotron photons are created
in the first pair creation burst in all fields, leading in turn
to many further pair creation events in further generations.  We
shall see in the next section that this does not necessarily
imply a greater efficiency of energy transfer from the primary
beam to the pair plasma, since the excess synchrotron photons
which do not pair create wind up with a large portion of the
energy budget.

\section{GENERAL RESULTS: MONOENERGETIC KERNELS}

In this section we present a representative subset of
the general results of our
monoenergetic simulations.  The set of runs shown span
cosine complements $\mu$ from 0.1 down to $10^{-12}$, stellar magnetic
fields  $B_*$ of 
$10^{11}, 10^{12},$ and $10^{13}$ G, and initial photon
energies $\vep$ from $300$ to $10^5$.

\subsection{Efficiency of Plasma and Photon Creation}

The efficiency of the cascade is important to global models of the
pulsar magnetosphere as well as to models of the radio and
high-energy photon
emission.  Our monoenergetic runs allow us to extract both number
and energy conversion efficiencies.  Since an arbitrary number of
starting photons seed each kernel run, these efficiencies are
defined per seed photon.

We begin with the pair plasma.  To describe its density,
we use the conversion ratio diagnostic $C^l(\vep;
\mu,B_*)$, defined as the number of pairs produced per parent
photon.  This can be roughly thought
of as the number of generations of leptons produced in the
cascade.  This quantity will also serve as a weighting function for
determining  plasma parameters of composite cascades, which begin
with a spectrum of initial photon energies. 

Table 1 shows $C^l$, the lepton creation efficiencies found in our
runs.  We see that
$C^l$ increases with $B_*$ and $\vep$.  This is as expected.  For low
$\vep$, $\mu$, and $B_*$ the seed photons' mean free path for pair
creation is larger than the scale height of the magnetosphere, so they
escape intact and then $C^l = 0$. For a fixed $B_*$, we find that $C^l$ is
primarily a function of the (maximum) pair creation parameter of the
initial photons.  This initial parameter, $\chi_i$, is defined using
equation (1) as $\chi_i = \chi(\vep,B_*,\thm(\mu))$, where $\thm
= \arccos (1 - \mu)$ is the largest angle of the injected photons.
$C^l$ increases with angle for large angles, but at small angles
it often saturates at some
value $C^l \lap 1.$  This occurs when the primary photons are
unable to pair produce until they propagate far enough outward
to increase their angle with $\bb$.
This $C^l \lap 1$ regime -- for low field values and small angles --
is roughly limited by  $\chi_i \lap 0.01.$  (We shall see later
that the lepton DFs for these low-$C^l$ runs
are independent of the opening cosine complement $\mu$). 
For larger $\chi_i$ values, we find that a 
dual power law $C^l \propto \thm^a \vep^b$ fits the $C^l$ data
nicely, with
different values of $a$ and $b$ for different values of $B_*$.
We find $(a,b) \approx (0.5,0.8)$ for $B_* = 10^{13}$ and
$(a,b) \approx (0.4,0.5)$ for the lower $B_*$ runs.  

We can form another diagnostic by dividing 
the final plasma energy by the initial photon energy, 
to define an efficiency of energy transfer, $E^l(\vep; \mu, B_*)$,
from primary photons into the leptons.
Table 1 also lists these values.  It is important to note that,
contrary to the assumptions made in some current models,  $E^l < 1$ is
commonly the case.  The missing energy corresponds to photons which
escape the 
magnetosphere. The parameters which determine the energy efficiency
are  $B_*$ and the cosine complement $\mu$. 
$B_*$ determines the overall range of efficiencies: $E^l \sim 0 -
.03$, $\sim 0 - .35$, and $\sim .40 - 1.0$ for $B_* = 10^{11},
10^{12},$ and $10^{13}$ G, respectively.  Synchrotron
losses thus weigh heavily in the polar cap's energy budget if the
magnetic field is weak. In this limit, the efficiency increases with
decreasing opening angle at first, but drops off again when  the angle  
becomes small enough for
many of the seed photons to escape without producing pairs.
Remarkably, the efficiency is highly insensitive to the initial
photon energy $\vep$ (except at the pair creation threshold where
$C^l \leq 1$, where lowering $\vep$ causes more seed
photons to escape).

We also find that $E^l$ is roughly anticorrelated with $C^l$ for a
fixed $B_*$, and peaks at small $\mu$ and large $\vep$.
That is, many pairs are created only when relatively little
of the available energy goes into making pairs.  This is understandable
since copious synchrotron radiation accounts for both effects,
but unfortunate if one's polar cap model requires
an efficient creation of dense plasma.
Interestingly, the lepton energy efficiency is relatively
insensitive to the initial photon energy.
Fixing other parameters, lowering $B_*$ always lowers $E^l$,
and lowers $C^l$ except at very large $\vep$ where $C^l$
peaks for $B_* = 10^{12}$ G.

The photons which escape are important both as a diagnostic and as a
direct observable.  To quantify the photon count, we calculate
the photon conversion ratio, $C^p(\vep;\mu,B_*)$, defined as 
the number of escaping photons with $\eps \geq 2$
divided by the initial number of seed photons.
Table 2 lists the values of $C^p$ of the simulations, which
ranged from zero to several hundred.  Several trends
are evident in the table.  Not surprisingly, $C^p$
always decreases with decreasing angle, since the first generation
of leptons produced must have large $\pperp$ at birth for
many synchrotron photons to be produced.
For a fixed angle, $C^p$ usually increases with increasing
$\vep$.  The few exceptions to this were all in the $10^{13}$ G
field, where some escaping primary photons were
more likely to create pairs as $\vep$ was increased, lowering
$C^p$.  Finally, we note that $C^p$ was a
maximum for the {\it intermediate} value of $B_* = 10^{12},$
holding the other parameters fixed.  This behavior is
attributable to the fact that we need both efficient pair creation
(which requires a large $B_*$) and copious synchrotron emission
by the created leptons (which happens for low $B_*$) to produce
many escaping synchrotron photons.

The efficiency of energy transfer into energetic
photons, $E^p$, is also presented in Table 2.  $E^p$ is defined as the
ratio of the energy in escaping photons with $\eps > 2$ to the seed
photon energy.  The difference between unity and the sum of $E^p$
and $E^l$ represents
the fractional energy that went into photons softer than
$\eps = 2$.  In the weakest field ($10^{11}$ G), most of
the primary photon energy remains in hard photons, even at
large $\vep$ and large $\mu$.  Also in the weak field,
$E^p$ increases with decreasing opening angle, since photons
are less likely to pair create at small angles.
In the strongest field ($10^{13}$ G),
almost none of the primary photon energy remains in hard photons.
Decreasing $\mu$ tends to decrease $E^l$ in strong fields, since
synchrotron emission disappears for small angles there.  In
our intermediate field ($10^{12}$ G), low-energy seed photons
mimic the weak-field trends for $E^l$, and high-energy seed
photons mimic the strong-field trends.

The cascade at $10^{11}$ G (in a dipole field) is in fact so
dominated by photons that following just those with the potential
to pair create rapidly ate up available computing resources.  For
that reason, the statistics we obtained on leptons in the
$10^{11}$ G field are relatively poor compared with those in the
stronger fields, and we did not include the $10^{11}$ G field
in our composite simulations of \S 4.

\subsection{Plasma Distribution Functions}

From a radio emission perspective, the shape of the final
lepton DFs is one of the most interesting
results of the cascade simulation.  The shape of the DF determines
the nature and propagation of wavemodes in the plasma 
and possible instabilities which may be converted into radio emission.

Although we generated too many DFs in our
parameter-space survey to present them all here, we can describe
the general behavior  concisely.  The DF shape is overwhelmingly
controlled by the opening cosine complement $\mu$, with secondary
dependence on magnetic field and seed photon energy.  The shape of the
DF tends to saturate at both large and small angles.  

\begin{figure}
\centerline{\psfig{figure=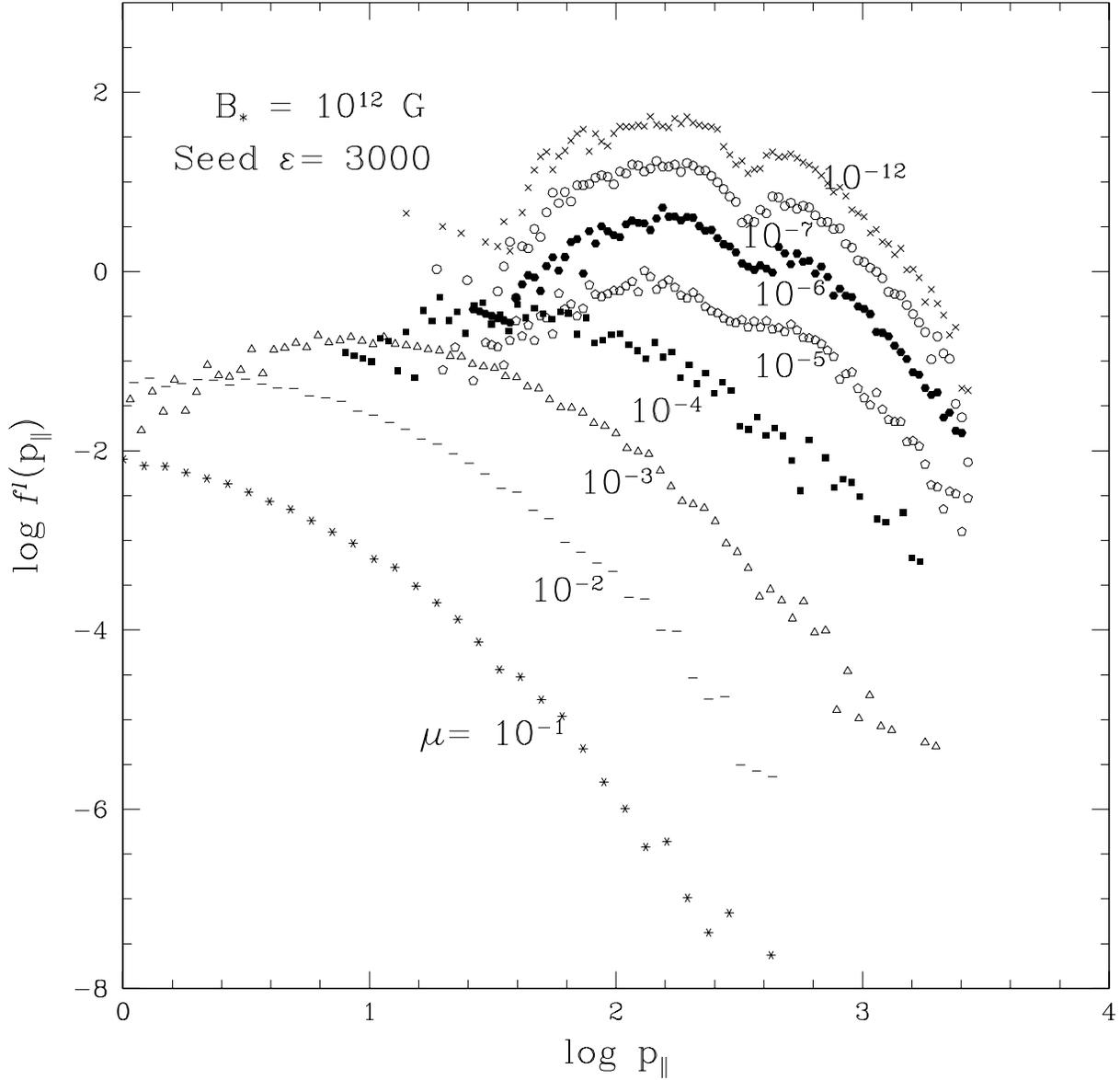,height=17cm,width=17cm}}
\caption[f3a.eps]  {(a) Lepton DF Dependence on Angle
Parameter $\mu$: $B_* = 10^{12}$ G.  Vertical offsets have been
made for clarity.  Notice
how the low-$\mu$ DFs all agree; the simulation results for low $\mu$
are in fact identical to the case $\mu = 0$.}
\end{figure}

\addtocounter{figure}{-1}

\begin{figure}
\centerline{\psfig{figure=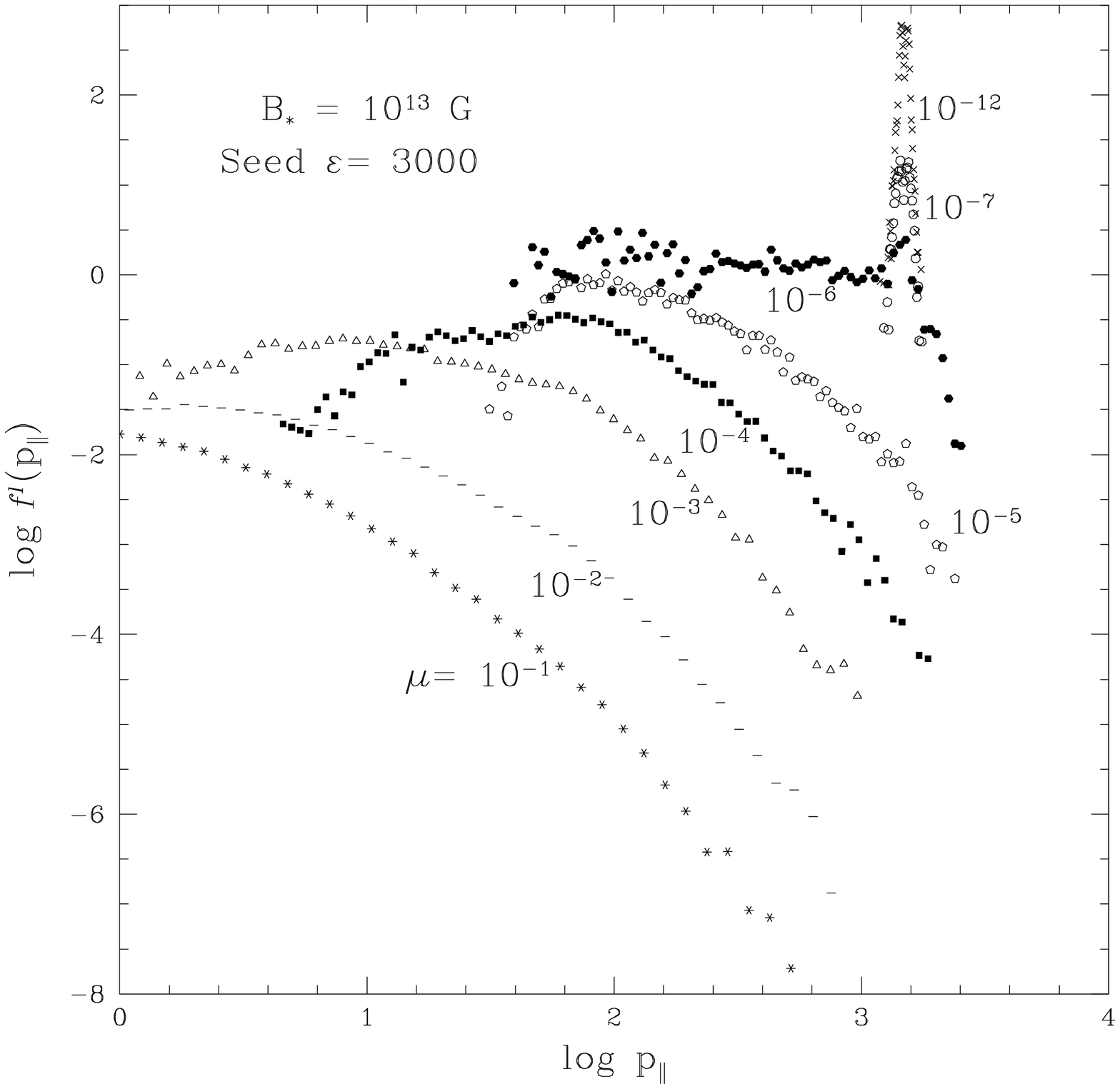,height=17cm,width=17cm}}
\caption[f3b.eps] {(b) Same as (a), for $B_* = 10^{13}$ G.  Note that the
low-$\mu$ DFs are much narrower than those in (a).
In contrast to (a), there is a slight $\mu$
dependence to the width of the low-$\mu$ DFs: narrower angles produce
narrower DFs.}
\end{figure}

We present a typical cross-section of the DFs for the entire
range of $\mu$ for fixed $\vep$ and two different $B_*$ values
in figure 3.  The saturated large-angle DF mentioned above
is demonstrated by the $\mu = 0.1$ DFs in the figure.
This DF is flat from $10^{-4} \leq \ppar \leq
1$, and drops off roughly as $\exp (-\ppar^{0.2})$ above this.  This
limiting large-angle shape is independent of magnetic field and seed
photon energy. As we decrease $\mu$, 
the DFs become narrower and peak at higher momentum.  At very
small angles (where $C^l$ is 1), these DFs again saturate to
a single shape, but now this shape depends on both $B_*$ and
$\vep$.  The double-humped shape of some of the low-angle
DFs in figure 3 is due to photons which give the members of
a created pair opposite $\ppar$ in the $\bk \cdot \bb = 0$
frame to conserve $\ppar$; the difference
is amplified by the boost to the star's reference frame.

As low angles are the relevant ones for seed photons in most models, 
we illustrate the dependence of low-angle DFs on $B_*$ and on $\vep$
in figure 4.  The important point illustrated here is 
that the width and other features in the shape are
determined by $B_*$, while the location of the peak is
determined by $\vep$.

\begin{figure}
\centerline{\psfig{figure=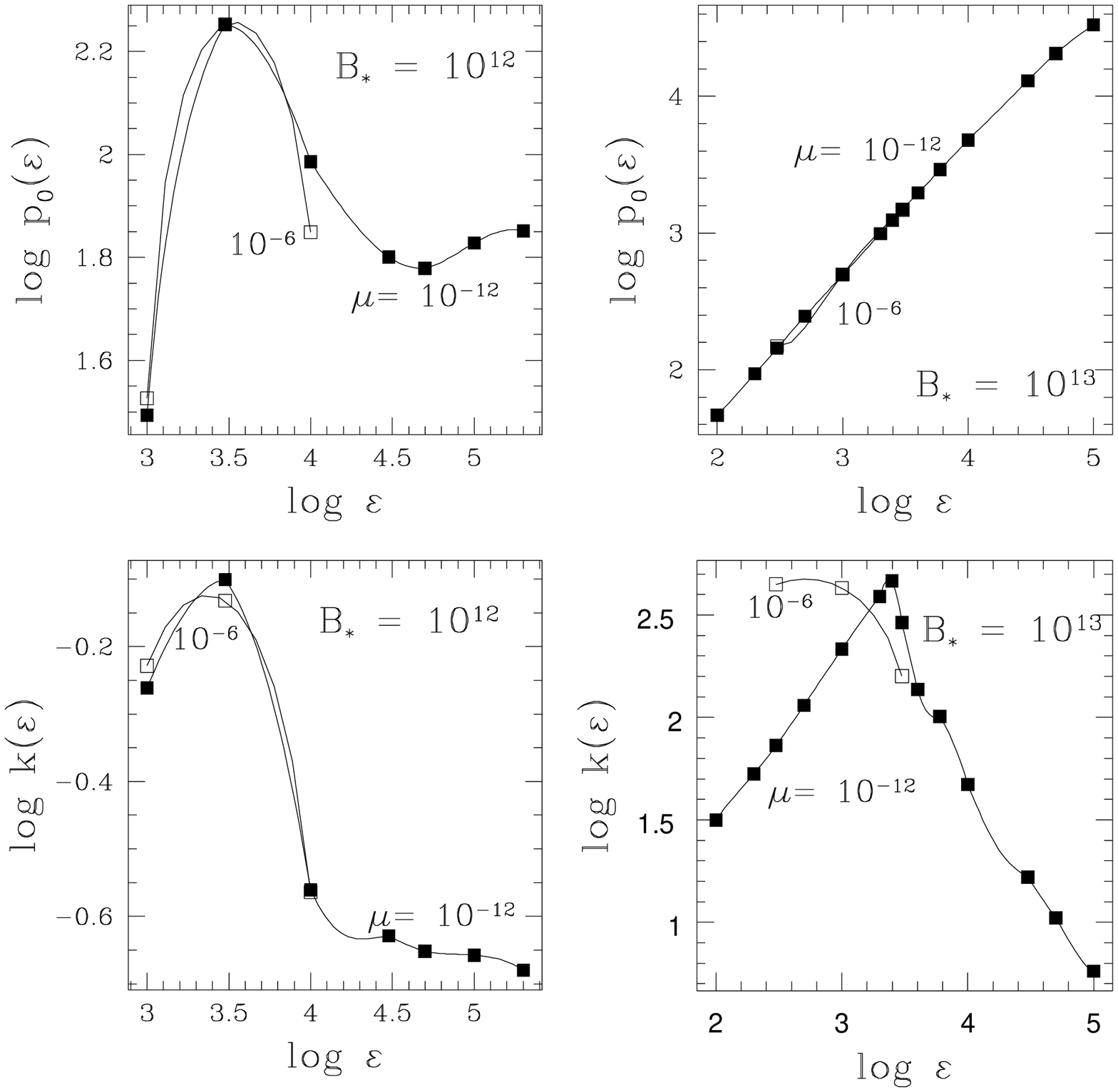,height=17cm,width=17cm}}
\caption[f4.eps]  {Lepton DFs for $\mu = 10^{-7}$.  The $B_* = 10^{12}$
and $10^{13}$ G DFs are shown for cosine complement $\mu = 10^{-7}$
and two values of seed photon energy.  A characteristic double-humped
shape is evident, and is due to unequal splitting of parent photon
energy to the lepton pair.  The much more narrow DFs in the $10^{13}$
G case give the much larger $K$ values seen in figure 5.  Vertical
offsets have been made for clarity.}
\end{figure}

Since we wished to use these lepton DFs from monoenergetic seed
photons to generate more realistic DFs (by convolution),
we determined an empirical expression which fits the monoenergetic DF
shapes.  We fit only the DFs for the $B_* = 10^{12}$ and
$10^{13}$ G runs, due to the poor plasma creation efficiencies in
the $10^{11}$ G runs.
We found that a parabola in $\log f(\ppar)$ -- $\log \ppar$
space is a remarkably good fit to most of the DFs.  We therefore used
the expression
$$
f(\ppar) = N \exp \biggl[ - K \biggl( \ln {{\ppar} \over {p_0}}
\biggr)^2 \biggr]
 \eqno(7)
$$
where $K$ and $p_0$ are fit parameters to be
determined for each `kernel' run.  This describes the fractional
number of particles $f(\ppar)$
having final momenta (along $\bb$) between $\ppar$ and
$\ppar + d\ppar$.  Thus, $f(\ppar)$ combines leptons from
all generations of the cascade.
The normalization
constant $N = \sqrt{K/\pi} \, e^{4 K} / p_0$ is chosen to ensure
that $\int_0^{\infty} f(\ppar) d\ppar = 1$ (this 
will be used in the composite simulations presented in \S 4).

We carried out
least-squares fitting to determine the optimal values of the 
parameters $p_0$ (location of the peak of $f$) and $K$ (roughly, the
inverse of the `width' of $f$).
The fits were empirically very good over
most of the parameter space, excepting the very largest
opening angles (which are not used in the composite runs).

\begin{figure}
\centerline{\psfig{figure=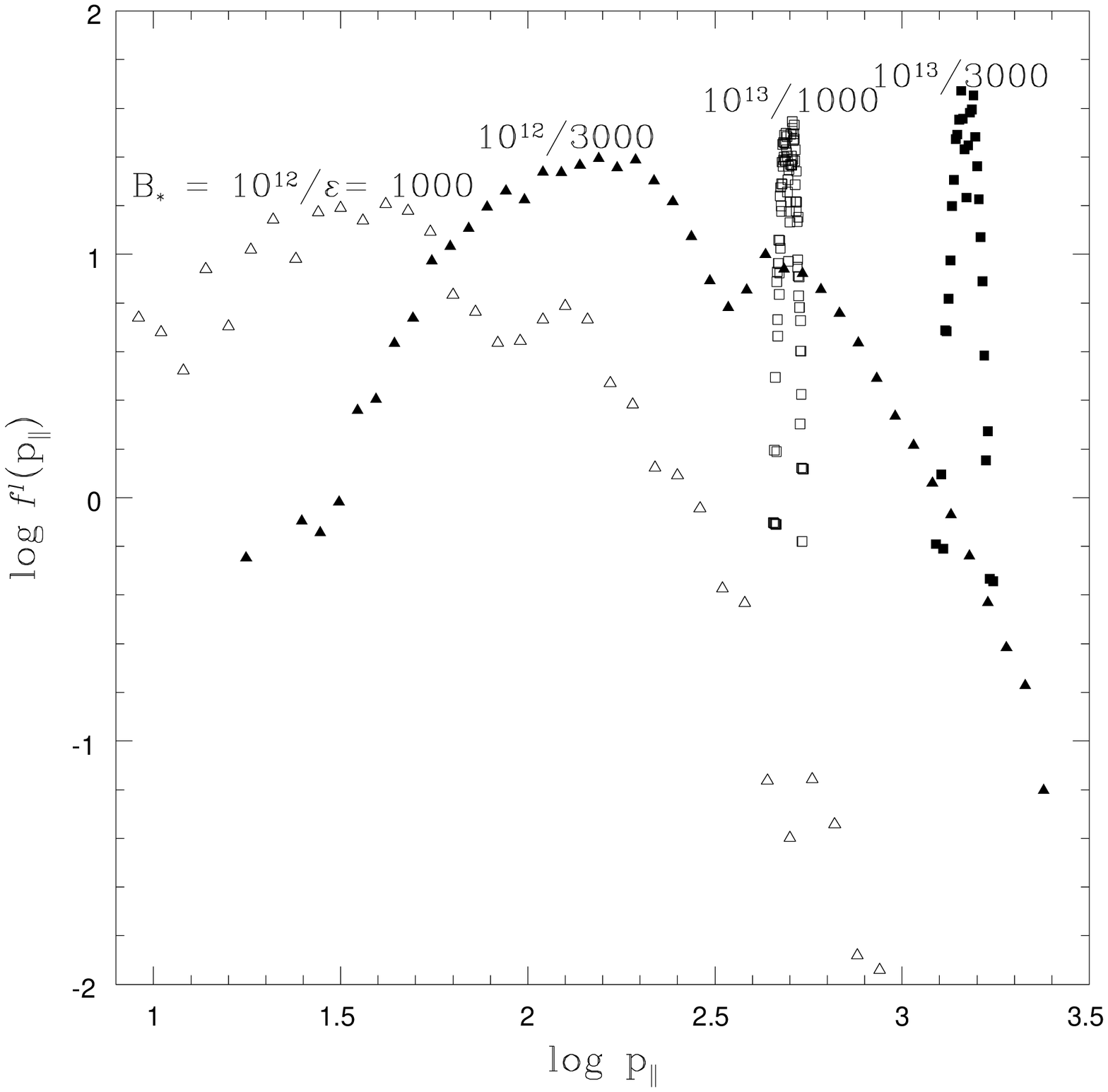,height=17cm,width=17cm}}
\caption[f5.eps]  {Lepton DF Fit Parameters.  Lepton 
DFs (viewed in the star's frame) were
least-square fit (in log space)
to a function of the form $f(\ppar) \propto \exp \bigl[ - K ( \ln \,
\ppar / p_0 )^2 \bigr] $.  The best-fit values of these 
two parameters are shown: $p_0$ is roughly the mean parallel momentum
of the DF, 
while $K$ is an inverse width.  Actual fits are shown as points, while
the lines show interpolations used for the composite DFs of \S
4.  Two values of $\mu$ are shown; the larger angle was only
considered for lower-energy seed photons.}
\end{figure}

As examples, the least-squared values for $p_0$ and $K$ for $\mu =
10^{-6}$ and $\mu = 10^{-12}$ are plotted in figure 5. The lines
connecting the fits represent interpolations used when 
simulating the composite DFs of \S 4. 
Values of $p_0$ ranged from below 1 to almost 200 for $B_* = 10^{12}$,
and were very nearly half the parent photon momentum for
$B_* = 10^{13}.$  Values of $K$ went from 0.1 to 0.8 for the
lower $B_*$, and were much
larger (10 to several hundred) for the higher $B_*$.
Note that the values of $K$
peaked at photon energies of a couple thousand,
then dropped at larger energies.  At large $\vep$, the photons
can produce pairs
with a larger spread of characteristics, which widens the DF.
Over the range where $K$ is
dropping, equation (7) was a poorer fit than elsewhere.
Also, at very large energies ($\vep > 10^5$) in the $10^{13}$ G
field, some pairs were born in excited Landau levels, and
synchrotron radiation then created further pairs (as in the
weaker-field cases).  However, these latter generations of leptons
were spread over a wide range of low $\ppar$ in the DF, and
equation (7) no longer described the DF shape well.  We did
not consider seed photon energies above ($\vep = 10^5$) in the
$10^{13}$ G field in \S 4 for this reason.

\subsection{Photon Spectra}

Distribution functions can be also be created for the escaping
photons.  To avoid confusion with the lepton DFs, we refer to
these as spectra throughout this work.  However, it is prudent
to emphasize that these represent the result of
binning together photon {\it number counts}, not energies. 

The photon spectra which come from our
simulations have fewer trends worth mentioning than the plasma DFs;
to first order, all are roughly the same shape.  The photon spectra are
all either  power laws which steepen at some $\eps_{\rm break}$, or
everywhere convex functions which may or may not exhibit
steepening at some $\eps_{\rm break}.$  The power laws have slopes
of $-.5$ to $-2$ below the break, and steepen to as much as $-4$ above
it.  The convex spectra have 
local slopes in the same ranges as the power laws.  The steeper
power laws
generally appear for the low field and large opening angles and
shallower power laws
for the high field and small opening angles.  For the $B_* = 
10^{12}$ runs, none of the photon spectra exhibited a true power
law over any significant range.

\begin{figure}
\centerline{\psfig{figure=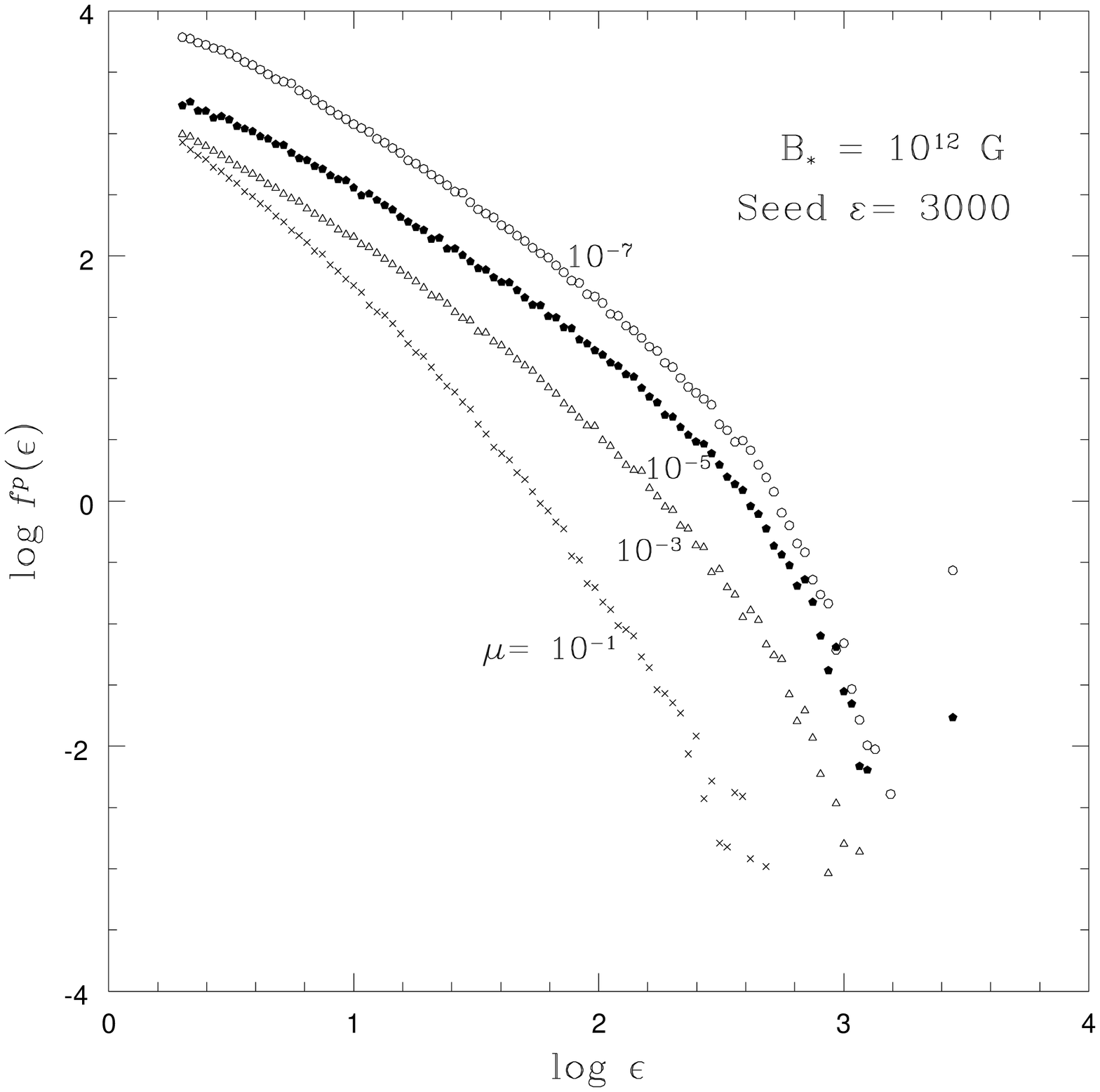,height=17cm,width=17cm}}
\caption[f6.eps]  {Photon Spectrum Dependence on $\mu$. A sample
of photon spectra is shown for $B_* = 10^{12}$ G and seed
photon energy $\vep = 3000$, for various cosine complements $\mu.$
Spectra for angles smaller than $\mu = 10^{-7}$ are identical to the
spectrum shown for $\mu = 10^{-7}$.
Vertical offsets have been made for clarity.  These typify
the spectral shapes found: a gradual steepening with energy
is seen, often
with a high-energy break above which the spectrum  sharply steepens.  The
points at $\eps = 3000$ are seed photons which did not create
pairs.}

\end{figure}

Samples of the convex photon spectra  for several values
of $\mu$ for a fixed $B_*$ and $\vep$ are shown in figure 6.
Except for small changes
in slope, the spectra all look remarkably similar.  Below the break,
if it occurs, the spectra are steeper for larger $\mu$.  This is
consistent with the results of Beskin \etal (1993),
who found that photon spectra steepened with each succeeding
generation of parent leptons (our largest angles produce the
largest $C^l$, which is roughly the number of lepton generations).
Above the break, the spectra are steeper for smaller $\mu$,
curving the spectra even more.  The bump at $\eps =
3000$ is just the primary photons which escaped the
magnetosphere without producing a lepton pair.

We can empirically (and somewhat arbitrarily) classify the
photon spectra according to type (power law or convex).
The power laws only seem to occur for the large field and
small angles (with a steep spectrum), and for the low field and large
angles (with a shallower spectrum).  A sample of the different
spectral types is given in figure 7, which
illustrates a sample power law with break and two convex
spectra with breaks.  It is evident that the distinction between
a power law and convex spectrum is somewhat fuzzy, as for
the $B_* = 10^{12}$ G case.  Note that these
spectra  differ from each other only by the parameter $B_*$.
In large fields, the break on the photon spectrum is due to pair
creation by the highest-energy synchrotron photons.  In low fields, 
the break is due to the intrinsic synchrotron spectrum
steepening.  The break locations ranged from
about $\eps \simeq 40 - 10^3$  for the lower two values of $B_*$,
increasing 
with increasing energy and generally with decreasing angle.
For $B_* = 10^{13}$, the break location was near $\eps = 100$ over
most of the parameter space, but decreased slightly for large
angles and low seed photon energies.
At smaller $\mu$, the synchrotron photons disappear in the
$B_* = 10^{13} G$ case, and thus so does the relevant
spectrum of figure 7.

\begin{figure}
\centerline{\psfig{figure=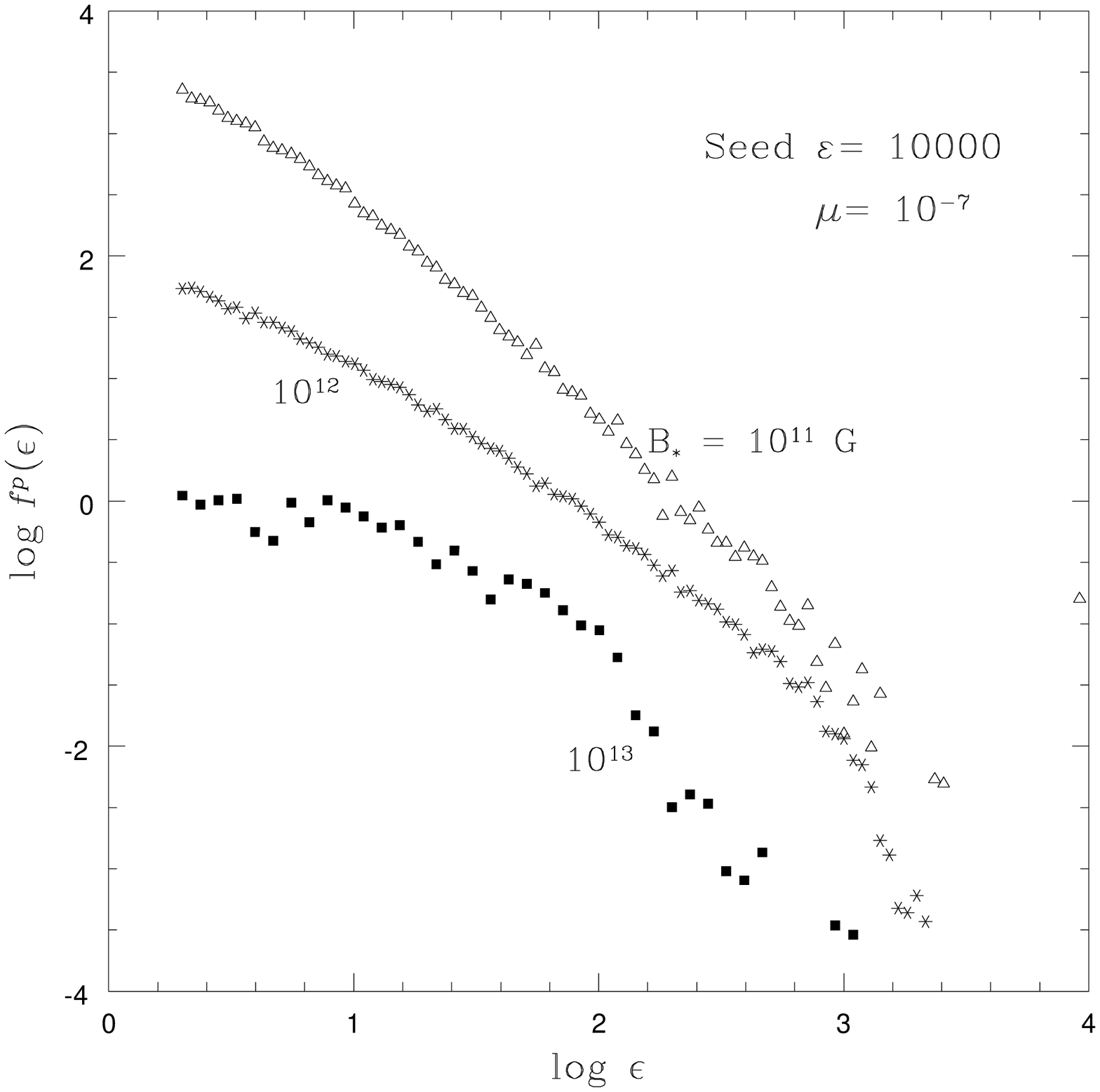,height=17cm,width=17cm}}
\caption[f7.eps]  {Photon Spectrum Dependence on $B_*$.  Seed
photon energy $\vep$
was $10^4$, and opening cosine complement $\mu$ was $10^{-7}$.
The high-energy breaks evident where the high-$B_*$ DFs steepen are
due to pair creation losses.  Statistics are
poor for the $B_* = 10^{13}$ G DF; at even smaller $\mu$ there are
no secondary photons in the strong field.  The DFs have been offset
vertically.}

\end{figure}

\section{APPLICATIONS TO PULSAR CASCADE MODELS}

Our monoenergetic runs gave us several useful quantities for cascades
deriving from monoenergetic photons.  
We found the distribution functions, $f^l(\ppar)$ and $f^p(\eps)$
(for leptons and photons, respectively).  We also found conversion
ratios $C^l(\vep)$ and $C^p(\vep)$, and energy efficiencies
$E^l(\vep)$ and $E^p(\vep)$.  
Cascades in pulsars are hardly likely to derive from
monoenergetic photons, however.  Many authors have assumed that the
photons which initiate the cascade are generated by curvature
radiation.  An interesting, more recent, alternative uses inverse
Compton scattering to seed the cascade. In this section we present
the results of using our monoenergetic results as `kernels' to
determine the cascades resulting from these two mechanisms.

We do this by convolving our results with the energy distribution of
seed photons which start the cascade, which we call $N(\vep)$.
We follow the standard picture in assuming that charges are pulled
from the star and accelerated to Lorentz factor $\gamma_b$ (these are
the `beam' charges).  We do not attempt to determine $\gamma_b$
self-consistently, but rather treat it as a parameter.  To create seed
photons, we consider both curvature radiation and magnetic
resonant inverse Compton scattering by the beam charges. 

For both of these processes,
the characteristic length over which a charge radiates most of its
energy is not too different from the scale of the polar cap and pair
formation region (Sturner 1995). We therefore take $N(\vep)$ to be the
composite photon spectrum (for the appropriate process) emitted by
a beam charge as it radiates most or all of its energy.  
The specific definitions of $N(\vep)$ are
given below in \S 4.1 and \S 4.2.

Once we know $N(\vep$), we can find
the composite photon and lepton distributions from
$$
F^p(\eps) = \int f^p(\eps;\vep) N(\vep) C^p(\vep) \, d \vep
\eqno(8)
$$
and
$$F^l(\ppar) = \int f^l(\ppar;\vep) N(\vep) C^l(\vep) \, d \vep \, .
\eqno(9)
$$
(Recall that $\vep$ refers to seed
photon energies, while $\eps$ is used for post-cascade photon
energies.)  
To evaluate the lepton DFs, we had to deal with the strongly varying
shapes of the monoenergetic DFs ($f^l(\ppar)$).  To do this, we used our
analytic approximations to the simulated DFs, described in \S 3.2,
and evaluated the integral in equation (9) numerically.  Parameters used
in the analytic approximations were interpolated for
values of $\vep$ between those actually simulated (shown by the
lines in fig. [5]).  As the monoenergetic photon spectra were simpler, we
dealt with them more simply. We binned the seed photons 
from a particular $N(\vep)$ around the $\vep$ values for which we
performed monoenergetic runs, and used these results to evaluate
the integral in equation (8) by discrete summation.

We computed number counts and energy efficiencies for the composite
runs in a similar fashion.  We define the multiplicity as the number
of pairs (half the number of leptons) created per primary
beam particle:
$$
M^l = \int F^l(\ppar) \, d\ppar \, = \int N(\vep) C^l(\vep) \,
d\vep \, . \eqno(10a)
$$

For photons, the multiplicity is the number of energetic photons
created per primary beam particle:
$$
M^p = \int_{\eps > 2} F^p(\eps) \, d\eps \, = \int N(\vep)
C^p(\vep) \, d\vep \, . 
\eqno(10b)
$$
We define fractional power as the ratio of energy escaping (in leptons
or in photons) to $\gamma_b$, the energy of a primary beam particle:
$$
P^l = {{1} \over {\gam_b}} \int F^l(\ppar) \gam(\ppar) \, d\ppar =
\int N(\vep) E^l(\vep) \, d\vep 
\eqno(11a)
$$
and
$$
P^p = {{1} \over {\gam_b}} \int F^p(\eps) \eps \, d\eps =
\int N(\vep) E^p(\vep) \, d\vep \, . 
\eqno(11b)
$$

In the composite runs we took $\thm \approx 1/\gam_b$, or
$\mu \sim 1/(2 \gam_b^2)$, for a particular value of primary
beam energy $\gam_b$.
We remind the reader that, although not stated
explicitly in the above expressions, all calculated quantities depend
upon the surface field strength, $B_*$.

\subsection{Curvature Radiation--Seeded Cascades} 

Curvature radiation (CR), arising from charges following curved field
lines, was the original mechanism proposed for generating
pair-creating photons in the pulsar magnetosphere (Sturrock 1971;
Ruderman \& Sutherland 1975), and is still commonly invoked today. 

The spectrum of photons
from a relativistic charge moving along a curved path
is formally identical to a synchrotron spectral shape,
but with the field lines' radius of curvature $\rho$
(and particle energy $\gam_b$)
determining the spectrum's characteristic energy $\vep_c = \hbar
\om_c /(m c^2) = 3 c \hbar \gamma_b^3 / (2 \rho \, m c^2) $.  
The instantaneous spectrum of curvature
radiation is thus easily found from the standard synchrotron power
formula, with the above substitution:
$$
{ d N (\vep ) \over d t }   = {{ \sqrt{3}} \over {2 \pi}}
{\alpha c \over \rho} { \gam_b \over { \vep}}
F\biggl( {{\vep} \over {\vep_c}} \biggr) 
\eqno(12)
$$
where $F(x) = x \int_x^\infty K_{5/3} (t) \, dt$, $K_{5/3}$ is a
modified Bessel function of the second kind, and $\alpha$ is
the electromagnetic fine-structure constant.  The function
$F(\vep/\vep_c)$ peaks at $0.29 \vep_c$, and falls off at higher
energies as $e^{-\vep / \vep_c}$.

The total spectrum emitted by a primary charge as it loses all of its
energy will be the integral of (12) over time, as $\gamma_b$ decays.
There is some model-dependent ambiguity in doing so, however:
the accelerating $\be$ field may still be present as
curvature radiation is happening, so $\gamma_b$ may be subject
to replenishment.  Furthermore, integrating equation (12) over
$\gamma_b$ is numerically somewhat tedious.
However, we note two useful approximations.  All of the
high-energy photons are emitted
early in the particle's life, before it has escaped the pair formation
region.  In addition, the high-energy end of the integrated spectrum
is very similar in shape to the high-energy end of the spectrum in
(12).

With these in mind, we may approximate the curvature spectrum
integrated over a short time by the instantaneous spectrum in
(12), normalized to the energy emitted over a short time.  For
overall normalization, we chose to let the beam particles radiate
over a time of 1 km/$c$, typical of acceleration
regions in polar cap models.  For all but one of our
composite curvature simulations, this is a short enough time
that only a small fraction of $\gamma_b$ is lost, and the
instantaneous shape of (12) is nearly identical to the
integrated spectrum accounting for decrementing $\gamma_b$ over
time accordingly.  (The exception is our energy-boosted $\gamma_b
= 5 \times 10^6$ run, for which beam replenishment is necessary
to keep radiation at constant power over this distance.)
We therefore used as a curvature seed spectrum
$$
N_{\rm CR} (\vep )  = {{ \sqrt{3}} \over {2 \pi}}
{(1 {\rm km}) \over \rho} { \gam_b \alpha \over { \vep}}
F\biggl( {{\vep} \over {\vep_c}} \biggr).
\eqno(13)
$$

Since this is a pair cascade simulation, we need to choose
curvature parameters $\gamma_b$ and $\rho$
that will lead to pair production.
(We remind the reader that our monoenergetic calculations treat a
simple dipolar magnetic field; the parameter $\rho$ does not need to
be specified there.)  If we
choose values typical of standard models of the polar flux tube ({\it
e.g.}, Cheng \& Ruderman 1977), we
find that CR cannot easily produce photons energetic enough to
generate a cascade.  We carried out simulations with $\rho = 100$ km
(an appropriate radius of curvature for magnetic dipole field lines),
and found that $\gam_b \gap 10^7$ is needed in order to
overcome the opacity limit and allow
significant pair creation.

Such high beam energies can be hard to
achieve in the standard models; pair cascades are difficult to
initiate in those models ({\it e.g.}, Weatherall \& Eilek 1997).  
Some authors have attempted to get around this problem by
invoking highly curved field configurations.  We wished to follow
this tradition, but without specifying a specific (unmotivated)
non-dipolar field geometry.  We therefore performed a second set of 
curvature spectrum seeded runs, boosting the characteristic curvature
energy $\vep_c$ by a factor of 10 by lowering $\rho$ to 10 km in
equation (13).  (Boosting curvature radiation in this manner is
frequently done in the literature to encourage a pair cascade;
however, we should emphasize that
we only boosted the photon spectrum and did not change
the field configuration.)  These curvature seeded runs are
referred to as the `energy-boosted' runs in what follows.

\subsection{Magnetic Resonant Inverse Compton--Seeded Cascades}

Inverse Compton scattering (ICS) of ambient thermal photons by a
primary particle beam has recently been gaining favor as a cascade
seed mechanism (\eg Daugherty \& Harding 1986; Kundt \& Schaaf
1993; Sturner \& Dermer
1994; Harding \& Muslimov 1998). In the strong pulsar magnetic field, the
scattering process happens with greatly increased cross-section
at resonances of the cyclotron frequency (Daugherty \& Harding
1986; Dermer 1990; Chang 1995).  If ICS is an important process in the
magnetosphere, radiation losses may limit primary beam energies
(to $\gam_b \sim 10^4$) while producing photons capable of pair
creation (Sturner 1995).  Indeed,
when ICS losses are significant, beam energies may well never
approach those necessary for energetic CR.

The lowest, and most important resonant
interaction (Daugherty \& Harding 1986) happens when the unscattered
photon frequency $\om$ equals the cyclotron
frequency $\om_B$ (or $\eps = \eps_B = \hbar \om_B /mc^2 = B/\bc$ )
in the
particle's rest frame.  Relativistic kinematics then gives for the
scattered photon energy
$ \vep_s = \eps_B / \gam_b (1 - \beta \cos \theta)$
in the stellar frame, where $\theta$ is the angle the initial
photon's direction makes with $\bb$.
The absolute limits on the scattered energy are therefore at the
extremes of $\cos \theta$: $\eps_B/(2 \gam_b) \leq \vep \leq
2 \gam_b \eps_B.$  When $\gam_b$ is large and all photon angles are
present, this represents a large frequency range into which
photons are scattered by magnetic resonant ICS.

The seed photon spectrum resulting from ICS depends upon the
ambient photons present (such as thermal photons above the polar
cap), and so cannot easily be
calculated analytically.  We rely, therefore, on numerical simulation
of magnetic ICS photon spectra from various 
source beams, from Daugherty \& Harding (1989; DH89).  Their
figure 9 shows the total spectrum for magnetic resonant ICS from a
monoenergetic primary beam.   It shows an approximate power-law of
slope $\sim -1.2$ in photon number versus energy.
The power-law seems to extend to very low
energies, and cuts off sharply at about $2 \gam_b \eps_B$, which is
the absolute limit derived above.  Although they
considered only a somewhat low-energy beam ($\gam_b = 200$), we
extrapolated their results to beams of higher energy without
modification.

We therefore assumed a magnetic ICS spectrum
$$ 
N_{\rm ICS}(\vep) = { {0.8 \gam_b m c^2} \over 
{\vep_{\rm max}^{0.8}}}  \vep^{- 1.2} , 0 < \vep <
\vep_{\rm max} 
\eqno(14)
$$
The cutoff energy  was related to the Lorentz factor of the primary
beam by $\vep_{\rm max} = 2 \gam_b \eps_B$. 
The normalization was chosen to give complete energy loss of
the primary beam into scattered photons.

\subsection{Results of Composite Simulations}

We computed composite CR-seeded cascades for primary beam
energies in the range $10^6 \leq \gam_b \leq 10^7$, and with seed
photons corresponding to two different curvature radii ($\rho = 100 $
km and $\rho = 10 $ km), corresponding to regular and `energy-boosted'
curvature spectra, respectively.  (As described above, a pure
dipolar field was used in all
simulations; our choices of $\rho$ are
significant only as a parameter in the curvature emission spectrum
they define.)  We also computed  ICS 
seeded-cascades for lower primary beam energies, in the range $10^3
\leq \gam_b \leq 5 \times 10^5$.  In all cases we computed cascades
for magnetic fields $B_* = 10^{12}$ and $10^{13}$G.
Very high beam energies for $B_* = 10^{13}$ were not simulated
since equation (7) no longer provided a good fit to the
necessary high-$\vep$ monoenergetic DFs there.

\subsubsection{Efficiencies and Multiplicities}

Some of the more important cascade quantities are
the fractions of the primary beam's power
which go into cascade-produced leptons and into energetic
photons, and the overall numbers of each created.  We present
these results in Tables 3 (for leptons) and 4 (for photons), and
recall that these quantities are defined per primary beam particle
(in contrast with the monoenergetic efficiencies, which were
defined per seed photon).  Formulae used to define and
calculate these quantities were provided in equations (10a) and (10b).

Table 3 summarizes the overall lepton number multiplicity $M^l$ and
lepton fractional power $P^l$ for all our composite runs.
We see that the lepton multiplicity and fractional power always
increase with beam energy.
The high-field case showed much better energy conversion than the
low-field case; this is due to both the higher probability
of pair creation and to the fewer synchrotron photons
created during cascades in strong fields.
For the $B_* = 10^{12}$ magnetic ICS computations, no cascade
develops for beam energies below $\gam_b = 2 \times 10^4$.
In general, lepton multiplicities were much higher for the most energetic
curvature-seeded cascades than for any ICS-seeded cascades,
but the ICS cascades generally
had better energy conversion efficiencies.  

Table 4 summarizes energetic
photon multiplicities and fractional power values
for all of our composite runs.  
The general results seen in the lepton table are also seen here,
although there are a few differences.  The photon multiplicity
$M^p$ generally increases with $\gam_b$, except for CR-seeded
cascades in the $10^{13}$ G field.  Photon fractional power
$P^p$ nearly always increases with $\gam_b$; again, the exceptions
are for CR in the $10^{13}$ G field.
Comparing Tables 3 and 4,
we also note that for $B_* = 10^{12}$ G, $M^p \gg M^l$, due to
the large number of lower-energy synchrotron photons generated.
At $B_* = 10^{13}$ G, $M^p < M^l$ for ICS-seeded and
high-energy CR-seeded cascades, due to the dearth of
secondary photons in these one-generation cascades.
At low primary beam energies, CR-seeded cascades produce many
photons which do not pair create, so $M^p$  is still large.
We note that the fractional powers do not generally add to
unity.  Any deficit is due to synchrotron photons with energies
$\eps < 2$.  The excess power in the energy-boosted $\gamma_b
= 5 \times 10^6$ run requires beam replenishment, and is due
to our choosing consistent normalization for $N_{\rm CR}$
throughout; see discussion preceding equation (13) for this.

\subsubsection{Lepton Distribution Functions}

One of the primary reasons for this work was to find the plasma DFs
predicted by pair cascade models.
Despite the enormous range in cascade parameters and pair creation
efficiencies encountered, 
we found that the composite DFs have remarkably similar shapes.
Figure 8 shows some sample 
curvature-seeded and ICS-seeded plasma DFs, sorted by
magnetic field strength and seed mechanism.  Normalizations
are as discussed above, except for the DF which was
multiplied up by the factor shown for visibility.
Only the energy-boosted curvature
composites are shown; the results for the standard curvature-seeded
runs
are qualitatively similar (but with lower efficiencies, as shown
in Table 3).  We see that both $B_*$ and the photon seed spectrum do
have some effect on the resultant plasma DFs.

\begin{figure}
\centerline{\psfig{figure=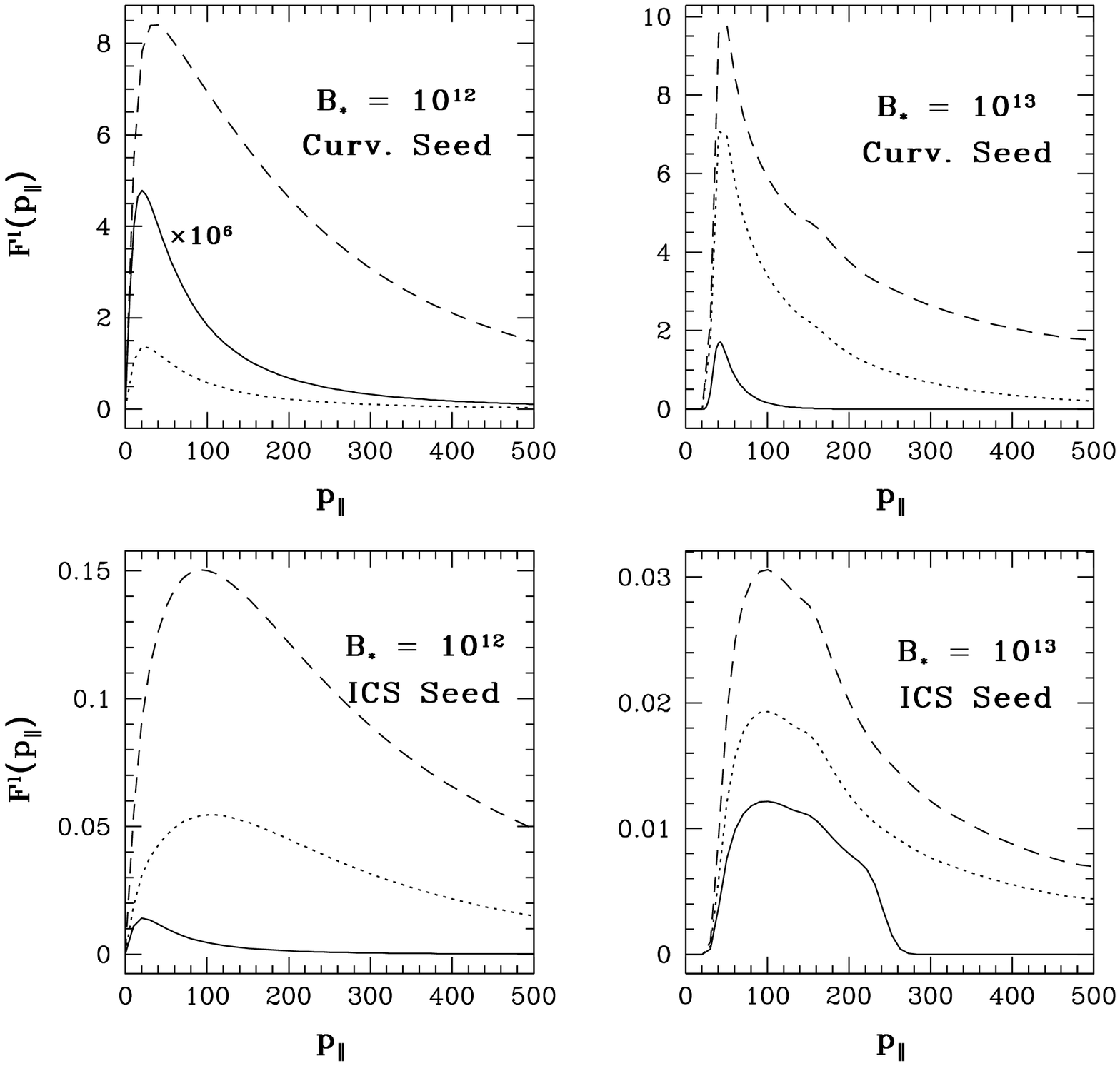,height=17cm,width=17cm}}
\caption[f8.eps]  {Composite Lepton DFs: Stellar Frame.
Sample DFs are given for pair
plasmas arising from both curvature and magnetic
resonant inverse Compton seed photon spectra.  The
DFs are shown as they appear in the (corotating) star frame.
The curvature-seeded runs shown here are all from the
`energy-boosted' runs, and beam energies of 1, 2
and 5 $\times 10^6 \, mc^2$, corresponding to the
solid, dashed, and dotted lines, respectively.
For the $B_* = 10^{12}$ ICS runs, the respective line types
are for beam energies of 2, 10, and 50 $\times 10^4$.
For the $B_* = 10^{13}$
ICS runs, beam energies are $10^3, 10^4,$ and $10^5$, respectively.
Normalizations are as discussed in \S 4 of the text.
Note the low-momentum cutoffs of the high-field DFs.  The
$B_* = 10^{12}$, $\gamma_b = 10^6$ curvature DF has been scaled
up by the factor shown for visibility.}
\end{figure}

For $B_* = 10^{13}$, the DFs all have a
low-$\ppar$ gap below about $20 mc$; {\it no}
slow leptons are present.  This is consistent with the narrower
`kernel' DF widths for $B_* = 10^{13}$ G, since low-energy
photons do not pair create at the small ($1/\gam_b$) angles
we considered here.  In contrast, the $B_* = 10^{12}$
DFs all extend down to $\ppar <  .01$. 
Synchrotron emission, significant in lower fields, helps to
achieve this in two ways.  First, leptons which
are born in excited Landau levels lose energy by synchrotron
radiation, and although $\beta$ is roughly conserved along $\bb$
the leptons still lose $\ppar$ 
as their total $\gamma$ decreases.  Second, synchrotron photons
which themselves create pairs are typically at lower energies
and larger angles with $\bb$ than the primary beam photons, so
the leptons are born with lower $\ppar$ to begin with (and may
still lose $\ppar$ by the first mechanism).

The curvature-seeded DFs
all peak somewhere between $\ppar = 10$ and $50$ for CR
cascades.  The low end has a very strong cutoff below
$\ppar = 10$, while the DF drops off slightly more gradually at
high $\ppar$.  The ICS-seeded DFs all peak at about $\ppar = 100$,
except for the $B_* = 10^{12}$, $\gam_b = 10^3$
DF which peaks at $\sim 20 mc$.  Generally, the ICS DFs drop
off more gradually than the CR DFs at large $\ppar$.
The $B_* = 10^{13}$ ICS DFs mimic the seed
spectrum power-law at high momenta (consistent with the
narrow monoenergetic `kernel' DFs for the strong field).

The results simplify if the DFs are transformed into the
center-of-momentum (CM) frame 
of the plasma.  To do this, we numerically found the average
Lorentz factor $\gave$ and average momentum $\pave$ of the
DF; then the transformation velocity is given by $\beta_{\rm CM}
= \pave/\gave$.  Once this is done, most of the DFs are found
to have shapes that are remarkably close to a {\it thermal}
distribution, which in this context means a 1-D relativistic
Maxwellian (J\"uttner - Synge) distribution:
$$
f(\ppar) = {{1} \over {2 K_1 (\zeta)}} \exp[ - \zeta \, \gam(\ppar)]
\eqno(15)
$$
where $\zeta = mc^2/k_B T$ is the inverse temperature of the
distribution, $K_1$ is the modified Bessel function, and
$\gam(\ppar) = \sqrt{1 + \ppar^2}.$

\begin{figure}
\centerline{\psfig{figure=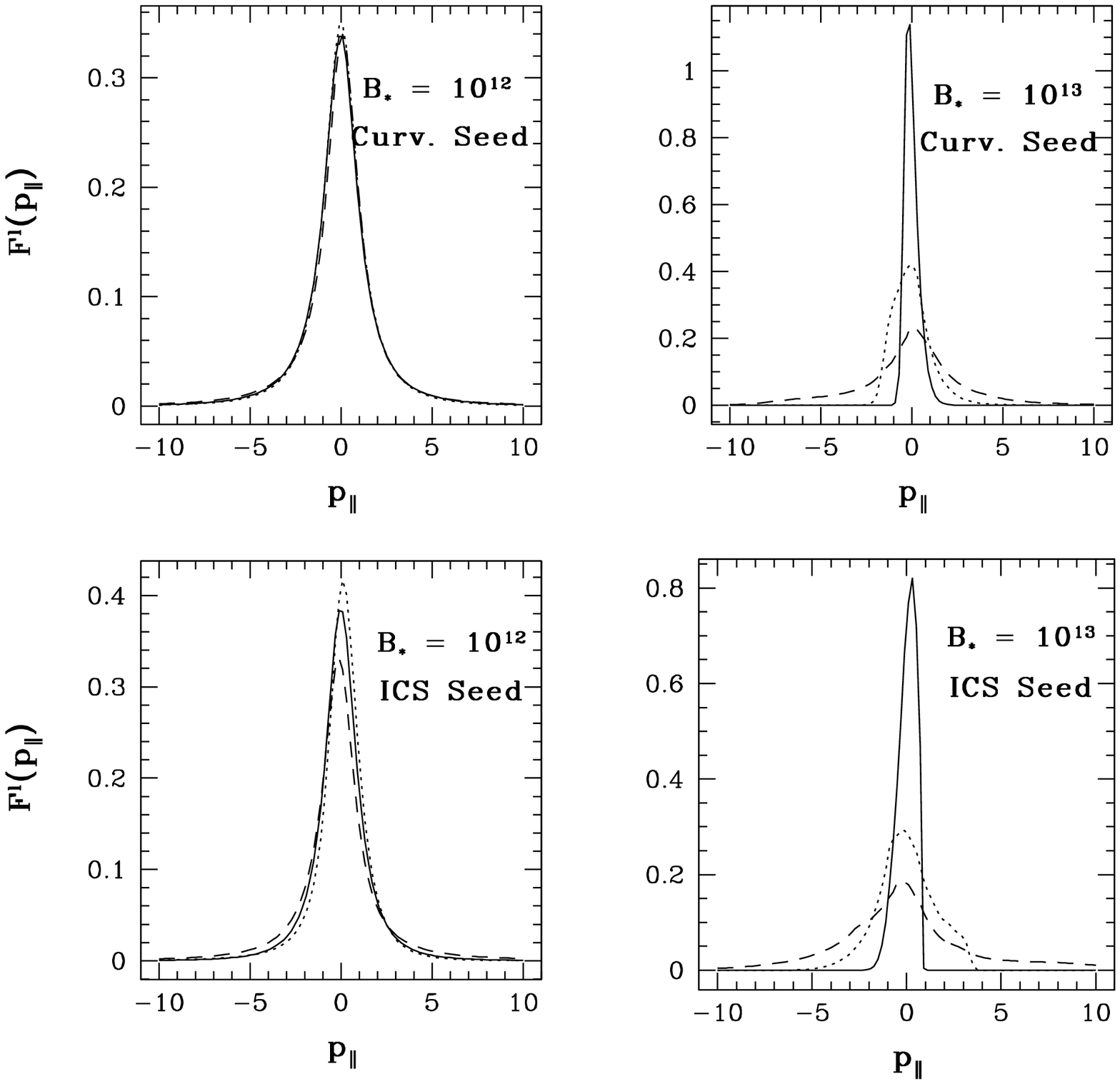,height=17cm,width=17cm}}
\caption[f9.eps]  {Composite Lepton DFs: CM frame.
The DFs have been transformed to their center-of-momentum
frames, and have been normalized to 1.  Beam energies are
associated to line types as in figure 8.
Statistics of the CM transformation and
Maxwellian fit parameters in the CM frame are given in Table 5.}
\end{figure}

Figure 9 shows the DFs in their respective
CM frames.  These DFs have all been
normalized to $\int \! f(\ppar) \, d\ppar = 1$.  The DFs for $B_* =
10^{12}$ G  are nearly indistinguishable from their corresponding
best fits (which we do not show) to equation (13), while the higher
field DFs show asymmetries not present in the Maxwellian fits.
Also, the DFs for $B_* = 10^{12}$  have nearly identical shapes
(temperatures) when seen in their CM frames, while the $B_* = 10^{13}$
DFs show temperature dependence with beam energy.

We present statistics of the DFs in Table 5.
Best-fit parameters of equation (15) are given, along with
the CM frame transformation Lorentz factors $\gcm = 
(1 - \betacm^2)^{-1/2}$.  Values of $\gcm$ ranged from 38 -- 890,
showing that the secondary plasma is indeed moving relativistically
outward.  However, values of the inverse temperature $\zeta$ show
that the plasma is only mildly relativistic in its own frame, with
$k_B T \sim mc^2$ for all the low-field cascades, and $k_B T \sim$
several $mc^2$ for high beam energies and cooler at lower beam energies
in the high-field cascades.  Even when $\gave$ was much larger than
$\gcm$, {\it the intrinsic plasma temperature is relatively low}, and the
apparent width of the DF in the stellar frame is due to the
Lorentz boost when passing to that frame (see also Weatherall 1994).

\subsubsection{Photon Spectra}

We also generated photon spectra for the composite simulations.  
Figure 10 shows the photon spectra for the energy-boosted
curvature seeded and the ICS seeded cascades.  The relative
heights of each spectrum correspond directly to the relative
normalizations for the cascade seeds.  The most important trend to
note is that the lower-field runs have steep photon spectra
for lower-energy \grys (for the largest $\gamma_b$ values),
while the high-field runs have flat spectra.  This directly
reflects the production, or not, of abundant synchrotron
photons in a multi-step cascade.

\begin{figure}
\centerline{\psfig{figure=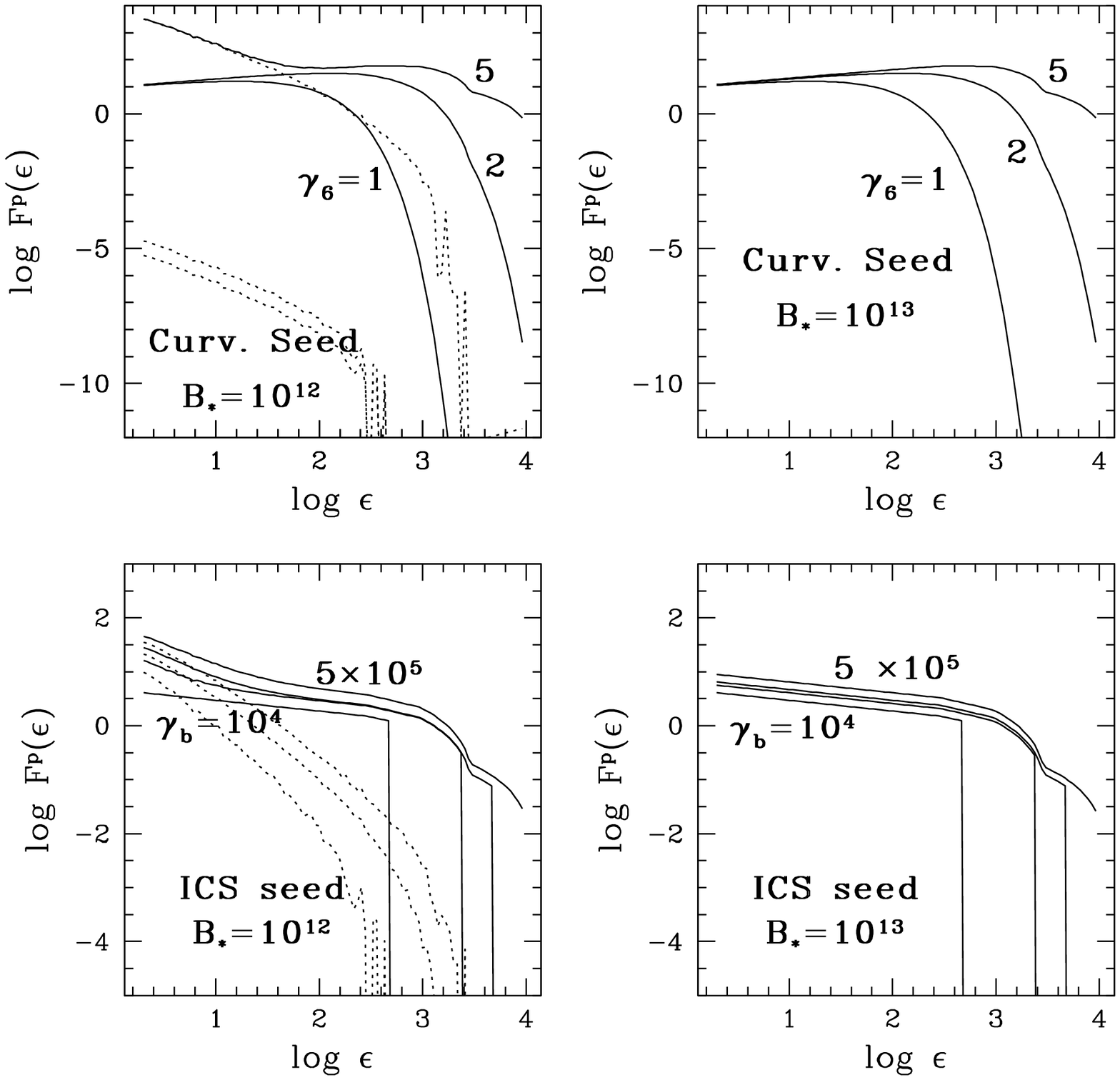,height=17cm,width=17cm}}
\caption[f10.eps]  {Composite Photon Spectra.  Spectra are given
for \grys of $\eps > 2$, from curvature and ICS seeded pair cascades. 
The curvature-seeded runs were from our set
with `energy-boosted' seed photons.
For the CR-seeded spectra, $\gamma_6$ refers to the primary beam Lorentz
factor divided by $10^6$.
Normalizations are as discussed in \S 4 of the text.  Both the leftover
seed photons and cascade-generated synchrotron photons are added
together for the spectra drawn with solid lines, while dotted lines are
the spectra from higher-generation synchrotron photons alone.  No
secondary photons are generated at these beam energies (assuming
$1/\gam_b$ beaming) in the $10^{13}$ G case, so only the seed
photons which did not produce pairs are shown.  Note that the
vertical axis for curvature cascades is twice the scale of the
ICS cascades.}
\end{figure}

The photon spectra for the $B_* = 10^{12}$ field include both
synchrotron photons generated in the cascade, and primary photons
which escaped the magnetosphere without producing a lepton pair.
The synchrotron contributions are shown separately (as the dotted
lines) alongside the total spectra in figure 10.
No secondary high-energy photons were generated
in the $B_* = 10^{13}$ runs that did not themselves pair produce,
so the only photons escaping from the cascade are lower-energy
seed photons which did not pair create.

For both CR and ICS cascades, at low beam energies, the photon spectra
are independent of $B_*$.  Each spectrum is
just the primary photon spectrum, depleted by those photons which
created pairs.  Synchrotron photons (which exist only for
the $B_* = 10^{12}$ G composites) are not numerous enough to
influence the photon spectral shape at low beam energies.
It is evident that the synchrotron photons decide the low-$\eps$
steep spectral shape at large beam
energies in the $10^{12}$ G field.  There is a high-energy break
above which the residual primary spectrum is seen again, due to the
steepness of the high-energy cutoff of the secondaries.

\section{DISCUSSION AND SUMMARY}

We have simulated the pair production cascade which may occur in
pulsar magnetospheres.  We did this by means of a numerical code which
tracks leptons and photons as they propagate upward above the magnetic 
polar cap of a rotating neutron star with a dipolar magnetic field.
The code simulates magnetic pair production by the photons and
quantized synchrotron emission by the leptons, beginning with a
population of monoenergetic seed photons.  We ran simulations with a
range of seed energies,
angles, and magnetic field strengths to obtain detailed results
over a substantial parameter range relevant to pair cascade models.
We used these results as kernels to model pair cascades produced by
an initial particle beam, either by curvature radiation or by
magnetic resonant inverse Compton scattering.

\subsection{Our Results}

Our main results can be summarized succinctly. 

\textbf{Cascade Onset and Development.}   We find that $B \sim
10^{11}$ G, or larger, is required in order for the cascade to go if
standard values are assumed for the primary beam energies.  This is a
consequence of the opacity condition, (6), and is {\it
independent of the magnetic field geometry}.
This suggests that slow rotation-powered
pulsars, and low-field millisecond pulsars, either have no pair
production or have seed photon energies much higher than the standard
models predict.  When the cascade occurs, we find that it develops and
ends spatially within $\delta r < r_*$ 
of the creation of the photon seeds, thus temporally within several
microseconds after each photon injection event.  In an actual pulsar
magnetosphere, it is possible that the seed photons are injected
continuously, but gap models with rapid temporal variability of the
accelerating potential and sparking (Ruderman \& Sutherland 1975) are
perhaps more likely due to the nonlinearity and magnitude of gap
accelerating potentials.  If sparking occurs, the cascade
timescale seen here may be directly connected to $\mu$sec flickering
known as microstructure.

\textbf{Efficiencies and Multiplicities.}  Our computed efficiencies
and multiplicities stand in contrast to  assumptions made in
the literature.  In particular, it is difficult to transfer
a large fraction of the primary beam's energy
to the pair plasma; some, often most, of it escapes as photons.
In addition, our computed multiplicities (number of pairs
produced per primary beam particle) are often small, only a few tens.
Only rarely do they reach the large values, $> 10^3$, which are
often assumed in the literature; curvature radiation with a large
beam energy seems to accomplish this best.

\textbf{The Pair Plasma.} We find that the plasma DFs in all cases
can be well described as relativistic Maxwellians {\it in the 
plasma's comoving frame}.  They generally have temperature
$k T \sim m c^2$.  The plasma flows out along field lines with Lorentz
factors  $\gamma_{\rm CM} \sim 100 - 1000$.  The resulting Lorentz boost
into the pulsar frame gives the characteristic, asymmetric shape seen
in  figure 8. 

\textbf{The Escaping Photons.}  We find that the escaping photon
spectra tend to be steep in lower magnetic fields, due to the
abundance of cooler, secondary synchrotron photons.  Conversely, they
tend to be flatter in higher fields, where synchrotron photons are
rare and the spectra are dominated by those seed photons which escaped
pair creation.

In addition to these general trends, the details of the cascade
development are sensitive to the initial field strength and also the
seed photon spectrum.

\textbf{Effect of Magnetic Field Strength.}
Qualitative differences in the pair cascade 
occur for initial magnetic field strengths of $B_* = 10^{13}$
compared to $B_* = 10^{12}$ G.  The `transition' field 
strength is likely near $B_* = 0.1 \bc \sim 4.4 \times 10^{12}$ G.
The differences are due to the abundance, or dearth, of secondary
synchrotron photons.  Below this transition field strength, leptons
tend to be born into excited Landau levels and copious synchrotron
emission takes place.  This occurs because the pair formation events
tend to be governed by opacity rather than energetics.  Above this
transition field, pair formation events are generally controlled by
energetics, so that leptons are usually born in their
lowest Landau state, and synchrotron emission is rare or absent.

\textbf{Effect of Seed Photon Source.} The overall cascade efficiency is 
sensitive to the seed mechanism.  We find that CR-seeded models
give pair multiplicities of up to a few times $10^3$, but typically
do not transfer much of the beam energy to the pair plasma.
In contrast, ICS-seed models give multiplicities of fewer
than $100$, but can efficiently transfer the beam energy into
the pair plasma (however, this requires ambient thermal photons
at the magnetic resonance of the beam, which we assumed at the outset).
Furthermore, the ICS mechanism operates at much lower beam
energies than does curvature radiation.

\textbf{Independence of Seed Photon Source.}
Despite the enormous range of cascade efficiencies, we find that the
plasma and photon distributions are fairly similar for both seed
photon  spectra which we used.
Perhaps the likenesses should not be too surprising, since
the general ingredients required for a pair creation cascade are
somewhat independent of seed mechanism.  First, photons must be
of sufficient energy to pair
produce; the lower-end shape of the photon spectrum does not
affect the plasma DFs at all. We chose photon seed spectra based on
typical beam energies in the literature; only a small fraction of the
seed photons these charges produce are able to seed the
cascade.  Second, a high-energy cutoff is always present in the
photon spectrum.  This cutoff, together with the low-energy
cutoff where the primary photons no longer produce pairs, limits
the effective (pair-producing) seed photon spectrum to a
relatively narrow range of energies.  Thus, many qualitative features
of the pair cascade are not very sensitive to the mechanism
of radiation which seeds the cascade (provided that it seeds
a cascade at all).

\subsection{Comparison to Other Simulations}

Other authors have performed numerical pair cascade simulations;
however, the focus has usually been on the final escaping
\grys rather than on the pair plasma.  

DH82 simulated cascades seeded by curvature radiation.   Our
CR-seeded photon spectra for high beam energies and field
strengths $B_* = 10^{12}$ G agree qualitatively with those
of DH82, but we find much softer spectra in the higher
$B_*$ case than they do.  The single lepton DF shown in DH82 (their
figure 7) is for $B_* = 10^{12}$
G, and a beam energy of $10^{13}$ eV ($\gam_b \sim 2 \times
10^7$).  It shows a rough power-law above $\ppar \sim 100 mc$,
and few pairs at lower momenta.  In addition, they predict that
{\it more} low-$\ppar$ pairs will be created in higher fields,
since photons of lower energy will be opaque to pair creation.
We believe that their conclusions (in contrast with what we present
here) can ultimately be traced to their assumption
that both members of the pair share the direction and half the energy
of the parent photon.  (Note
that DH82 predates Daugherty \& Harding 1983,
where they present energy-differential pair production
cross-sections.  In stronger fields, the lack of synchrotron
radiation is more significant than the ability of lower-energy
photons to pair create, and so the DFs still have fewer leptons
at low $\ppar$.)

Sturner \etal (1995)
investigated cascades seeded by inverse Compton radiation.
They do not present the final plasma DFs at all.  Our ICS-seeded
photon spectra for $B_* = 10^{12}$ agree qualtitatively with their
results for $B_* = 4 \times 10^{12}$.  Note that we find much flatter
spectra for our $B_* = 10^{13}$ simulations (for which Sturner \etal
did not offer spectra).

\subsection{Comparison to other Plasma DFs}

The plasma DF is  critical in determining the wavemodes the
plasma can support.  Knowledge of the DF is therefore crucial to
calculations of radio emission and signal propagation in the pulsar
magnetosphere.  Previous work has had to assume some some analytic
expression, not necessarily physically motivated.  Some work has
chosen analytically convenient 
forms, such as cold plasma (delta function), boxcar and bell-curve 
DFs, and relativistic Maxwellians (both in the comoving and pulsar
frames).  Our work supports one of these assumptions, namely, a 
comoving Maxwellian. Our final DFs are only marginally relativistically
hot, with temperatures $k_B T \sim mc^2$ in most cases, and lower
temperatures in $B_* = 10^{13}$ fields when the beam energy is
just above the threshold to seed a pair cascade.  The large spread
in $\ppar$ in the pulsar-frame DFs is simply due to of Lorentz 
boosting.

In addition, some authors have used an analytic representation of a DF
which Arons (1981) presented as a cartoon model.
The Arons cartoon-DF has a peak at $\ppar \sim$ a few $mc$. It drops
off exponentially above $\ppar \sim 10^3$--$10^4$. 
Both of these features are roughly consistent with our lower-field
results, where synchrotron photons are produced.  Our higher-field
results have a qualitatively similar shape but at higher $\ppar$
values. In addition, the Arons cartoon-DF 
has a flat region between these extremes.
We do not see this feature in our DFs, even in the $B_* = 10^{12}$ case.

\subsection{Impact on Future Work}

We anticipate that the results presented here
will be most relevant to plasma-based
modeling of the magnetosphere and of the radio emission mechanisms. 
We give two examples.

The pair plasma plays a critical role in the electrodynamics of the
polar flux tube;  it is often assumed to short out the parallel
electric field and terminate the acceleration region.  Ruderman \&
Sutherland (1975) estimated that 
the pair multiplicity should be around $\gam_b/\gam_{\rm pairs}$.
This is typically two orders of magnitude larger than the
multiplicities we find (except for the high-field magnetic ICS runs).
Shibata \etal (1998) agree with Ruderman \&
Sutherland that these large multiplicities
($10^3$--$10^5$) are necessary to short out the accelerating
$\be_\parallel$ field.  Our results suggest that the situation is not
so simple.   If complete shorting-out cannot be maintained, we might
expect a complex, non-steady plasma flow in the region.

Instabilities in the pair plasma are often assumed to give rise to
coherent radio emission.  One such mechanism is a two-stream
instability, driven by relative motion of the two signs of charge
({\it e.g.}, Buschauer \& Benford 1976).  The  high temperatures
assumed by some authors tend to suppress this instability
({\it e.g.}, Weatherall 1994);  our results however find a lower
temperature which may be favorable for this mechanism. We are
presently investigating the details of this instability in the pair
plasmas found in our calculations (Arendt \& Weatherall, in progress).

\section*{ACKNOWLEDGEMENTS}
We would like to thank Galen Gisler, who helped formulate the basic
calculation and set up the initial
code.  Many helpful discussions were had with members of the New
Mexico Tech pulsar group, including Tim Hankins, Jim Weatherall,
David Moffett, Jeff Kern, and Tracey DeLaney.  J.E. would also like
to thank Jon Arons and Don Melrose for useful discussions.  An
anonymous referee provided some useful suggestions, as well.
We are grateful for partial funding that was provided by NSF grants
AST-9315285 and AST-9618408.

\section*{APPENDIX: CODE DETAILS}

We present here some of the relevant physical and computational
details of our pair cascade code.  We discuss the kinematics
of the pair production process, and its implementation in our
code.  We also present the relevant details of synchrotron photon
production.

The probability for photon pair production can be described by
an attenuation coefficient (the inverse of the mean free path)
$R(\chi)$, where $\chi$ was defined above equation (2).
The (spin and polarization averaged)
differential attenuation coefficient is given (in the
frame moving along $\bb$ satisfying $\bk \cdot \bb = 0$) in Daugherty
\& Harding (1983) as
$$
{{dR(\kap,\chi)} \over {d\kap}} \approx {{2 \alpha} \over
{ \sqrt{27} \lambda_C}} {{B} \over {\bc}} {{2 + \kap (1-\kap)} \over
{ \chi \kap (1-\kap)}}
K_{2/3} \biggl[ {{1} \over {3 \chi \kap (1-\kap)} }\biggr] 
\eqno(A1)
$$
where $\alpha$ is the fine structure constant, $\lambda_C = \hbar/mc$,
$\kap$ is the fraction of photon energy
given to one member of the created pair, and $K_{2/3}$ is the
modified Bessel function.  The behavior of $dR/d\kap$ with $\kap$
determines how a given photon distributes its
excess (above $\eps = 2$) energy between the electron and positron.

We created a grid of numerically determined values of $dR/d\kap$
over a range of $\chi$ and $\kap$, and used interpolation when
needing $dR/d\kap$ in the code.
To find the total attenuation coefficient $R(\chi),$ we numerically
integrated $dR/d\kap$ over $\kap$ for each fixed value of $\chi$.  The
total $R(\chi)$ was then used to determine when to create a pair.
At each timestep, we multiplied the distance traveled by each photon
by the $R(\chi)$ appropriate to the photon's environment, and added
this `fractional mean free path'
to the photon's optical depth $\tau$, approximating
$$
\tau = \int R(\chi) \, ds.
\eqno(A2)
$$
After $\tau > 1$, we
created a lepton pair in place of the photon.  To find the split in
energy amongst the members of the pair, we needed partial
probabilities as well.  We integrated $dR/d\kap$
from $0$ to $\kap$ for a set of values of $\kap$ and normalized by
$R(\chi)$.  This function allowed us to use a random number to
choose (with interpolation) $\kap$ for a given pair production event.

Energy and momentum along $\bb$ are conserved in one-photon magnetic
pair production.  Momentum of the particles and photons across $\bb$ 
is not conserved; the magnetic field provides for the difference,
which allows the process
to occur kinematically.  Once we determined the fractional
energy $\kap$ given to one lepton, we chose a random number uniform
in $\cos^2 \theta$ (over the kinematically allowed angles) to
determine its pitch angle (in the $\bk \cdot \bb = 0$ frame).
The other lepton's energy and pitch angle were then simply
determined by the conservation laws.

We also required an expression for the synchrotron emissivity
of an electron or positron.  We needed  both the differential
emissivity as a function of emitted photon energy (to decide
how energetic a photon to make), and total synchrotron emissivity
over the interesting photon range ($\eps > 2 $).
Harding \& Preece 1987
give a useful approximation to the differential synchrotron
emissivity of a lepton of energy $\gam$, in a frame moving along
$\bb$ such that the lepton's momentum parallel to $\bb$ vanishes.
We divided this by the photon energy $\eps$ to get the differential
photon number production rate in a magnetic field of strength $B$:
$$
{{d \dot N(\gam,\eps,B)} \over {d \eps}} = {{\sqrt{3}} \over {2 \pi}}
\alpha \omega_B {{f} \over {\eps}} \biggl[ y \int_y^\infty K_{5/3}(x)
\, dx + y^3 f \, \biggl( {{3 \gam B} \over {2 \bc} } \biggr)^2
K_{2/3}(y) \biggr],
 \eqno(A3)
$$
where $y \equiv 2 \eps \bc / (3 f \gam^2 B),$
and $f \equiv 1 - \eps/\gam$ is the fraction of lepton energy
remaining after the photon emission.

The code created a synchrotron photon at the lepton's location
when the lepton's total energy would have
been lost if it radiated its energy away {\it linearly} at its present
rate, defining a sort of `optical depth' for the lepton's
momentum transverse to the field.   We then decrement this opacity
by the fraction of actual energy radiated in the photon creation
event, and continue the process until no more photons with $\eps > 2$
can be created by synchrotron radiation.  After that, the classical
synchrotron power formula is used on the lepton to decrement its
energy with time.  This approximate method was agrees
well with the classical synchrotron loss rate.

Once the decision is made to create a synchrotron photon, we use a
grid of $d \dot N /d \eps$ integrated from $0$ to $\eps$ on a set
of parameter values to determine the photon's energy, using a
similar algorithm to the one described above for pair production.

\vfill \eject

\begin{deluxetable}{ccccccc}

\tablecolumns{7}

\tablecaption{MONOENERGETIC CASCADE EFFICIENCIES: LEPTONS}


\tablehead{
\colhead{}
& \multicolumn{2}{c}{$ B_* = 10^{11}$ G}
& \multicolumn{2}{c}{$ B_* = 10^{12}$ G}
& \multicolumn{2}{c}{$ B_* = 10^{13}$ G} \\
\colhead{Cosine Complement $\mu$}
& \colhead{$C^l$} & \colhead{$E^l$}
& \colhead{$C^l$} & \colhead{$E^l$}
& \colhead{$C^l$} & \colhead{$E^l$}
}

\startdata

\multicolumn{7}{c}{$\vep = 300$} \\
$ 10^{-1}$
& .83   & .029
& 3.8   & .14
& 10    & .41
\\
$ 10^{-3}$
& \cd\tablenotemark{a} & \cd
& .81   & .23
& 1.5   & .48
\\
$ 10^{-5}$
& \cd & \cd
& \cd & \cd
& .91 & .91
\\
$ 10^{-7}$
& \cd & \cd
& \cd & \cd
& .91 & .91
\\
$ 10^{-12}$
& \cd & \cd
& \cd & \cd
& .91 & .91
\\

\multicolumn{7}{c}{$\vep = 1000$} \\
$ 10^{-1}$
& 1.2 & .015
& 11. & .14
& 27. & .40
\\
$ 10^{-3}$
& \cd & \cd
& 1.2 & .13
& 4.4 & .52
\\
$ 10^{-5}$
& \cd & \cd
& .27 & .076
& 1.1 & .75
\\
$ 10^{-7}$
& \cd & \cd
& .26 & .062
& .99 & .99
\\
$ 10^{-12}$
& \cd & \cd
& .26 & .064
& .99 & .99
\\

\multicolumn{7}{c}{$\vep = 3000$} \\
$ 10^{-1}$
& 3.8 & .016
& 32. & .14
& 73. & .41
\\
$ 10^{-3}$
& .75 & .023
& 3.5 & .13
& 12. & .52
\\
$ 10^{-5}$
& \cd & \cd
& .99 & .28
& 3.5 & .72
\\
$ 10^{-7}$
& \cd & \cd
& .92 & .29
& 1.0 & 1.0
\\
$ 10^{-12}$
& \cd & \cd
& .92 & .28
& 1.0 & 1.0
\\

\multicolumn{7}{c}{$\vep = 10000$} \\
$ 10^{-1}$
& 11. & .014
& 89. & .14
& 200 & .44
\\
$ 10^{-3}$
& 1.2 & .013
& 11. & .14
& 35. & .57
\\
$ 10^{-5}$
& .16 & .004
& 2.1 & .18
& 11. & .77
\\
$ 10^{-7}$
& .10 & .002
& 1.4 & .34
& 1.5 & .78
\\
$ 10^{-12}$
& .11 & .002
& 1.0 & .24
& 1.0 & .99
\\

\multicolumn{7}{c}{$\vep = 30000$} \\
$ 10^{-12}$
& .87 & .026
& 3.6 & .36
& 1.0 & .96
\\

\multicolumn{7}{c}{$\vep = 100000$} \\
$ 10^{-12}$
& 1.2 & .031
& 11. & .38
& 1.0 & .89
\\

\enddata

\label{crtab}
\tablenotetext{a} {No pair creation at these angles.}

\end{deluxetable}



\begin{deluxetable}{ccccccc}

\tablecolumns{7}

\tablecaption{MONOENERGETIC CASCADE EFFICIENCIES: PHOTONS}


\tablehead{
\colhead{}
& \multicolumn{2}{c}{$B_* = 10^{11}$ G} 
& \multicolumn{2}{c}{$B_* = 10^{12}$ G}
& \multicolumn{2}{c}{$B_* = 10^{13}$ G}
\\
\colhead{Cosine Complement $\mu$}
& \colhead{$C^p$} & \colhead{$E^p$}
& \colhead{$C^p$} & \colhead{$E^p$}
& \colhead{$C^p$} & \colhead{$E^p$}
}

\startdata

\multicolumn{7}{c}{$\vep = 300$}  \\
$10^{-1}$
& 390 & .998
& 24  & .569
& 15  & .366
\\
 $ 10^{-3}$
& \cd\tablenotemark{a} & \cd
& 10  & .537
& 5.0 & .322
\\
 $10^{-5}$
& \cd  & \cd
& \cd  & \cd
& .092 & .091 
\\
 $10^{-7}$
& \cd  & \cd
& \cd  & \cd
& .091 & .090
\\
 $10^{-12}$
& \cd  & \cd
& \cd  & \cd
& .098 & .098
\\

\multicolumn{7}{c}{$\vep = 1000$} \\
$ 10^{-1}$
& 55  & .833
& 73  & .580
& 42  & .353
\\
$ 10^{-3}$
& \cd & \cd
& 35  & .722
& 15  & .322
\\
 $ 10^{-5}$
& \cd & \cd
& 9.2 & .855
& 6.2 & .030
\\
 $ 10^{-7}$
& \cd  & \cd
& 11   & .876
& .015 & .014
\\
$ 10^{-12}$
& \cd  & \cd
& 10   & .873
& .012 & .012
\\

\multicolumn{7}{c}{$\vep = 3000$} \\
$10^{-1}$
& 150  & .840
& 210  & .593
& 110  & .346
\\
 $ 10^{-3}$
& 110  & .905
& 87   & .737
& 38   & .292
\\
 $ 10^{-5}$
& \cd  & \cd
& 32   & .485
& 5.5  & .081
\\
 $ 10^{-7}$
& \cd & \cd
& 32  & .476
& 0   & 0
\\
$ 10^{-12}$
& \cd & \cd
& 33  & .486
& 0   & 0
\\

\multicolumn{7}{c}{$\vep = 10000$} \\
 $ 10^{-1}$
& 470 & .849
& 630 & .614
& 320 & .386
\\
 $ 10^{-3}$
& 220 & .948
& 260 & .730
& 100 & .241
\\
 $ 10^{-5}$
& 59 & .989
& 96 & .668
& 15 & .069
\\
 $ 10^{-7}$
& 39 & .993
& 52 & .403
& .38 & .002
\\
$ 10^{-12}$
& 42  & .992
& 51  & .409
& 0   & 0
\\

\multicolumn{7}{c}{$\vep = 30000$} \\
$ 10^{-12}$
& 370 & .944
& 140 & .350
& 0   & 0 
\\

\multicolumn{7}{c}{$\vep = 100000$} \\
$ 10^{-12}$
& 580 & .937
& 420 & .298
& 0   & 0 
\\

\enddata
\label{photcrtab}
\tablenotetext{a} {No pair creation at these angles.}

\end{deluxetable}



\begin{deluxetable}{ccccc}

\tablecolumns{5}

\tablecaption{COMPOSITE CASCADE EFFICIENCIES: LEPTONS}


\tablehead{
\colhead{}
& \multicolumn{2}{c}{$ B_* = 10^{12}$ G}
& \multicolumn{2}{c}{$ B_* = 10^{13}$ G} \\
\colhead{$\gamma_{\rm beam}$}
& \colhead{$M^l$} & \colhead{$P^l$}
& \colhead{$M^l$} & \colhead{$P^l$}
\\
}

\startdata

\multicolumn{5}{c}{Curvature Seed} \\

$1 \times 10^6$
& $ 0 $ & $ 0 $
& $ 5.6 \times 10^{-7}$ & $ 4.5 \times 10^{-11}$
\\
$2 \times 10^6 $
& $7.5 \times 10^{-6}$ & $ 2.3 \times 10^{-11}$
& 1.7 & $1.2 \times 10^{-4}$
\\
$5 \times 10^6 $
& 18 & $1.5 \times 10^{-3}$
& 160 & .020
\\
$ 1 \times 10^7$
& 360 & .053
& 660 & .19
\\

\\
\multicolumn{5}{c}{Energy-Boosted Curvature Seed} \\

$ 1 \times 10^6$
& $4.8 \times 10^{-4}$ & $5.7 \times 10^{-9}$
& 15 & $2.5 \times 10^{-3}$
\\
$2 \times 10^6$
& 29 & $4.5 \times 10^{-3}$
& 500 & .12
\\
$5 \times 10^6 $
& 2200 & .72
& 3400  & 2.3\tablenotemark{a}
\\

\\
\multicolumn{5}{c}{Magnetic ICS Seed} \\
$1 \times 10^3$
& \cd\tablenotemark{b} & \cd
& 2.0     & .55
\\
$2 \times 10^3$
& \cd    &  \cd
& 3.5      & .73
\\
$5 \times 10^3$
& \cd    &  \cd
& 5.8      & .87
\\
$1 \times 10^4$
& 0.   & 0.
& 7.8     & .93
\\
$2 \times 10^4$
& 1.3     & $ 8.4 \times 10^{-3}$
& 10     & .96
\\
$5 \times 10^4$
& 9.9     & .092
& 14      & .96
\\
$1 \times 10^5$
& 22     & .18
& 17     & .96
\\
$2 \times 10^5$
& 40     & .25
& \cd\tablenotemark{c}    & \cd
\\
$5 \times 10^5$
& 80     & .30
& \cd    & \cd
\\

\enddata

\label{comptab}
\tablenotetext{a}{Replenishment is assumed in this case, to
keep consistency with the normalization used for the other curvature
runs; see discussion preceding equation (13) in the text.}
\tablenotetext{b}{No cascade at these energies.}
\tablenotetext{c}{Simulations not performed; lepton DF fit
poor at high energies.}

\end{deluxetable}


\begin{deluxetable}{ccccc}

\tablecolumns{5}

\tablecaption{COMPOSITE CASCADE EFFICIENCIES: PHOTONS}


\tablehead{
\colhead{}
& \multicolumn{2}{c}{$ B = 10^{12}$ G}
& \multicolumn{2}{c}{$ B = 10^{13}$ G} \\
\colhead{$\gamma_{\rm beam}$}
& \colhead{$M^p$}
& \colhead{$P^p$}
& \colhead{$M^p$}
& \colhead{$P^p$}
}

\startdata

\multicolumn{5}{c}{Curvature Seed} \\
$1 \times 10^6$
& 26 & $ 1.5 \times 10^{-4} $
& 26 & $ 1.5 \times 10^{-4} $
\\
$2 \times 10^6 $
& 1200 & $ 1.5 \times 10^{-3}$
& 1200 & $ 1.3 \times 10^{-3}$
\\
$5 \times 10^6 $
& $1.3 \times 10^4$ & .020
& 2800 & $3.1 \times 10^{-3}$
\\
$ 1 \times 10^7$
& $9.5 \times 10^5$ & .095
& 320 & $2.3 \times 10^{-3}$
\\

\\
\multicolumn{5}{c}{Energy-Boosted Curvature Seed} \\

$ 1 \times 10^6$
& 6500 & .019
& 6500 & .016
\\
$2 \times 10^6$
& $3.3 \times 10^4 $ & .14
& 1200 & .030
\\
$5 \times 10^6 $
& $ 9.1 \times 10^5$ & 1.2\tablenotemark{a}
& 1500 & .022
\\

\\
\multicolumn{5}{c}{Magnetic ICS Seed} \\
$1 \times 10^3$
& \cd\tablenotemark{b} & \cd
& .32     & .009
\\
$2 \times 10^3$
& \cd    &  \cd
& .36      & .005
\\
$5 \times 10^3$
& \cd    &  \cd
& .46      & .003
\\
$1 \times 10^4$
& 32   & .97
& .53     & .002
\\
$2 \times 10^4$
& 37    & .56
& .61     & .001
\\
$5 \times 10^4$
& 270    & .56
& .73    & $ 4.0 \times 10^{-4}$
\\
$1 \times 10^5$
& 1000     & .35
& \cd\tablenotemark{c}     & \cd
\\
$2 \times 10^5$
& 1200    & .20
& \cd    & \cd
\\
$5 \times 10^5$
& 2600   & .099
& \cd    & \cd
\\

\enddata

\label{compphottab}
\tablenotetext{a}{Replenishment is assumed in this case, to
keep consistency with the normalization used for the other curvature
runs; see discussion preceding equation (13) in the text.}
\tablenotetext{b}{No cascade at these energies.}
\tablenotetext{c}{Simulations not performed; lepton DF fit
poor at high energies.}

\end{deluxetable}


\begin{deluxetable}{ccccccc}

\tablecolumns{7}

\tablecaption{COMPOSITE CASCADES: LEPTON DF STATISTICS}


\tablehead{
\colhead{}
& \multicolumn{3}{c}{$ B_* = 10^{12}$ G}
& \multicolumn{3}{c}{$ B_* = 10^{13}$ G}
\\
\cline{2-4} \cline{5-7}\\
\colhead{$\gamma_{\rm beam}$}
& \colhead{$\bar \gamma$ \tablenotemark{a}}
& \colhead{$\gamma_{\rm CM}$ \tablenotemark{b}}
& \colhead{$\zeta$ \tablenotemark{c}}
& \colhead{$\bar \gamma $}
& \colhead{$\gamma_{\rm CM}$}
& \colhead{$\zeta$}
}

\startdata

\multicolumn{7}{c}{Curvature Seed} \\

$1 \times 10^6 $
& \cd\tablenotemark{d} & \cd & \cd
& 39  & 38  & 37
\\
$2 \times 10^6 $
& 160 & 80 & 1.1
& 55 & 52 & 9.8
\\
$5 \times 10^6 $
& 170 & 92 & 1.1
& 350 & 160 & 1.0
\\
$ 1 \times 10^7 $
& 700 & 330 & 1.3
& \cd  & \cd & \cd
\\

\\
\multicolumn{7}{c}{Energy-Boosted Curvature Seed} \\

$ 1 \times 10^6 $
& 160 & 79 & 1.1
& 59 & 55 & 7.8
\\
$2 \times 10^6 $
& 150 & 81 & 1.2
& 170 & 130 & 1.6
\\
$5 \times 10^6 $
& 460 & 210 & 1.1
& 1500 & 590 & .60
\\

\\
\multicolumn{7}{c}{Magnetic ICS Seed} \\
$1 \times 10^3$
& \cd & \cd & \cd
& 140 & 120 & 4.5
\\
$2 \times 10^3$
& \cd &  \cd & \cd
& 210 & 170 & 2.5
\\
$5 \times 10^3$
& \cd &  \cd & \cd
& 370 & 260 & 1.3
\\
$1 \times 10^4$
& \cd  & \cd & \cd
& 590 & 350 & .94
\\
$2 \times 10^4$
& 110 & 62 & 1.3
& 920 & 470 & .72
\\
$5 \times 10^4$
& 270 & 160 & 1.3
& 1700 & 890 & .53
\\
$1 \times 10^5$
& 370 & 220 & 1.5
& 2700 & 890 & .44
\\
$2 \times 10^5$
& 510 & 270 & 1.4
& \cd\tablenotemark{e}  & \cd & \cd
\\
$5 \times 10^5$
& 980 & 400 & 1.0
& \cd  & \cd & \cd
\\

\enddata

\label{stattab}

\tablenotetext{a}{Average Lorentz factor of this DF.}
\tablenotetext{b}{Lorentz factor of transformation to
center-of-momentum frame.}
\tablenotetext{c}{Best fit relativistic Maxwellian inverse temperature
(found in CM frame): $\zeta = m c^2/k_B T$.}
\tablenotetext{d}{No cascade at these energies.}
\tablenotetext{e}{Simulations not performed; lepton DF fit
poor at high energies.}

\end{deluxetable}


\begin{references}

\reference{as79} Arons, J., \& Scharlemann, E. T. 1979, \apj, 231, 854

\reference{ar81} Arons, J. 1981, in Proc. Varenna Summer School
and Workshop on Plasma Astrophysics, ed. T. D.
Guyenne \& G. L\'evy (Noordwijk: ESA), 273

\reference{ar83} Arons, J. 1983, \apj, 266, 215

\reference{ab86} Arons, J., \& Barnard, J. J. 1986, \apj, 302, 120

\reference{ar92} Arons, J. 1992, in IAU Colloquium No. 128,
ed. T. H. Hankins, J. M. Rankin, \& J. A. Gil
(Zielona G\'ora, Poland: Pedagogical Univ. Press), 56


\reference{bgi88} Beskin, V. S., Gurevich, A. V., \& Istomin, Ya. N.
1988, Astrophys. Space Sci., 146, 205

\reference{bgi93} Beskin, V. S., Gurevich, A. V., \& Istomin, Ya. N.
1993, Physics of the Pulsar Magnetosphere (Cambridge: CUP)

\reference{busben77} Buschauer, R. \& Benford, G. 1977, MNRAS, 179, 99

\reference{bam86} Bussard, R. W., Alexander, S. B., \& M\'esz\'aros,
P. 1986, Phys. Rev. D, 34, 440

\reference{chang95} Chang, H.-K. 1995, Astron. Astrophys., 301, 456

\reference{cr77} Cheng, A. F., \& Ruderman, M. A. 1977, \apj, 214, 598

\reference{dh82} Daugherty, J. K., \& Harding, A. K. 1982, \apj,
252, 337

\reference{dh83} Daugherty, J. K., \& Harding, A. K. 1983, \apj,
273, 761

\reference{dh86} Daugherty, J. K., \& Harding, A. K. 1986, \apj,
309, 362

\reference{dh89} Daugherty, J. K., \& Harding, A. K. 1989, \apj,
336, 861

\reference{dl75} Daugherty, J. K., \& Lerche, I. 1975, \ApSS, 38, 437

\reference{derm90} Dermer, C. D. 1990, \apj, 360, 197


\reference{gmg98} Gedalin, M., Melrose, D. B., \& Gruman, E. 1998,
Phys. Rev. E, 57, 3399

\reference{} Goldreich, P. \& Julian, W. H., 1969, \apj, 157, 869


\reference{ha96} Hankins, T. H., in Pulsars:  Problems and Progress
(IAU Colloquium 160), ed. S. Johnston, M. A. Walker and M. Bailes (San
Francisco:  ASP), 197 

\reference{hp87} Harding, A. K., \& Muslimov, A. G., 1998, \apj, 508, 328

\reference{hp87} Harding, A. K., \& Preece, R. 1987, \apj, 319, 939

\reference{} Hirotani, K. \& Shibata, S., 1999, \mnras, 308, 54

\reference{ja75} Jackson, J. D. 1975, Classical Electrodynamics
(New York: John Wiley \& Sons)

\reference{} Johnson, M. H., \& Lippman, B. A., 1949,
Phys. Rev., 76, 828

\reference{kazetal91} Kazbegi, A. Z., Machabeli, G. Z., \& Melikidze,
G. I. 1991, MNRAS, 253, 377

\reference{ks93} Kundt, W. \& Schaaf, R. 1993, Astrophys. Space Sci.,
200, 251

\reference{lyub96} Lyubarskii, Yu. E. 1996, Astron. Astrophys., 308, 809

\reference{lbm99} Lyutikov, M., Blandford, R. D., \& Machabeli, G.
1999, MNRAS, 305, 338

\reference{mel92} Melrose, D. B. 1992, in IAU Colloquium No. 128,
ed. T. H. Hankins \etal
(Zielona G\'ora, Poland: Pedagogical Univ. Press), 307

\reference{mel95} Melrose, D. B. 1995, J. Astrophys. Astr., 16,
137

\reference{rank83} Rankin, J. M. 1983, \apj, 274, 333

\reference{} Romani, R. W., 1996, \apj, 470, 469

\reference{} Rudak, B. \& Dyks, J., 1999, \mnras, 303, 477

\reference{rs75} Ruderman, M., \& Sutherland, P. 1975, \apj, 196, 51

\reference{smt98} Shibata, S., Miyazaki, J., \& Takahara, F. 1998,
MNRAS, 295, L53

\reference{sturn95} Sturner, S. J. 1995, \apj, 446, 292

\reference{ds94} Sturner, S. J. \& Dermer, C. D.,  1994, \apj, 420, L79

\reference{stm95} Sturner, S. J., Dermer, C. D., \& Michel, F. C.
1995, \apj, 445, 736

\reference{sturr71} Sturrock, P. A. 1971, \apj, 164, 529


\reference{te74} Tsai, W., \& Erber, T. 1974, Phys. Rev. D, 10, 492

\reference{usmel96} Usov, V. V., \& Melrose, D. B. 1996, \apj, 464,
306

\reference{jw94} Weatherall, J. C. 1994, \apj, 428, 261

\reference{} Weatherall, J. C. \& Eilek, J. A.,  1997, \apj, 474, 407


\reference{} Zhang, B. \& Harding, A. K., 2000, \apj, 532, 1150

\end{references}
\end{document}